\documentclass[twocolumn]{aastex63}

\newcommand{\ri}{$R_{\rm in}$}

\newcommand{\tmax}{$T_{\rm max}$}

\newcommand{\romanone}{\uppercase\expandafter{\romannumeral1}}

\usepackage{graphicx}
\usepackage[caption=false]{subfig}
\usepackage{xcolor}
\usepackage{url}
\usepackage{booktabs}
\usepackage{amsmath}
\usepackage{hyperref}

\newcounter{column_number}
\shortauthors{Liu et al.}
\shorttitle{Parameter Space of FUors}

\begin{document}

\defcitealias{connelley18}{CR18}

\title{Diagnosing FUor-like Sources: \\
The Parameter Space of Viscously Heated Disks in the Optical and Near-IR}



\correspondingauthor{Hanpu Liu, Gregory Herczeg}
\email{liuhanpu@stu.pku.edu.cn, gherczeg1@gmail.com}

\author[0000-0003-2488-4667]{Hanpu Liu}
\affil{Kavli Institute for Astronomy and Astrophysics, Peking University, Yiheyuan 5, Haidian Qu, 100871 Beijing, China}
\affiliation{Department of Astronomy, Peking University, Yiheyuan 5, Haidian Qu, 100871 Beijing, China}

\author[0000-0002-7154-6065]{Gregory J. Herczeg}
\affil{Kavli Institute for Astronomy and Astrophysics, Peking University, Yiheyuan 5, Haidian Qu, 100871 Beijing, China}
\affiliation{Department of Astronomy, Peking University, Yiheyuan 5, Haidian Qu, 100871 Beijing, China}

\author[0000-0002-6773-459X]{Doug Johnstone}
\affiliation{NRC Herzberg Astronomy and Astrophysics,
5071 West Saanich Road,
Victoria, BC, V9E 2E7, Canada}
\affiliation{Department of Physics and Astronomy, University of Victoria,
3800 Finnerty Road, Elliot Building,
Victoria, BC, V8P 5C2, Canada}

\author[0000-0003-1894-1880]{Carlos Contreras-Pe\~na}
\affiliation{Department of Physics and Astronomy, Seoul National University, 1 Gwanak-ro, Gwanak-gu, Seoul 08826, Republic of Korea}
\affiliation{School of Space Research, Kyung Hee University, 1732, Deogyeong-daero, Giheung-gu, Yongin-si, Gyeonggi-do 17104, Republic of Korea}

\author[0000-0003-3119-2087]{Jeong-Eun Lee}
\affiliation{School of Space Research, Kyung Hee University, 1732, Deogyeong-daero, Giheung-gu, Yongin-si, Gyeonggi-do 17104, Republic of Korea}

\author[0000-0002-8537-6669]{Haifeng Yang}
\affil{Kavli Institute for Astronomy and Astrophysics, Peking University, Yiheyuan 5, Haidian Qu, 100871 Beijing, China}
\affiliation{Department of Astronomy, Peking University, Yiheyuan 5, Haidian Qu, 100871 Beijing, China}

\author{Xingyu Zhou}
\affil{Kavli Institute for Astronomy and Astrophysics, Peking University, Yiheyuan 5, Haidian Qu, 100871 Beijing, China}
\affiliation{Department of Astronomy, Peking University, Yiheyuan 5, Haidian Qu, 100871 Beijing, China}

\author{Sung-Yong Yoon}
\affiliation{School of Space Research, Kyung Hee University, 1732, Deogyeong-daero, Giheung-gu, Yongin-si, Gyeonggi-do 17104, Republic of Korea}
\affiliation{Korea Astronomy and Space Science Institute, 776, Daedeok-daero, Yuseong-gu, Daejeon, 34055, Republic of Korea}

\author{Ho-Gyu Lee}
\affiliation{Korea Astronomy and Space Science Institute, 776, Daedeok-daero, Yuseong-gu, Daejeon, 34055, Republic of Korea}

\author[0000-0002-1932-3358]{Masanobu Kunitomo}
\affiliation{Department of Physics, Kurume University, 67 Asahimachi, Kurume, Fukuoka, 830-0011, Japan}

\author{Jessy Jose}
\affiliation{Indian Institute of Science Education and Research (IISER) Tirupati, Rami Reddy Nagar, Karakambadi Road, Mangalam (P.O.), Tirupati 517 507, India}

\begin{abstract} 
FU Ori type objects (FUors) are decades-long outbursts of accretion onto young stars that are strong enough to viscously heat disks so that the disk outshines the central star.
We construct models for FUor objects by calculating emission components from a 
steady-state viscous accretion disk, a passively-heated dusty disk, 
magnetospheric accretion columns, and the stellar photosphere. 
We explore the parameter space of the accretion rate $\dot{M}$ and stellar mass $M_*$ to investigate implications on the optical and near-infrared spectral energy distribution and spectral lines.
The models are validated by fitting to multi-wavelength photometry of three confirmed FUor objects, FU Ori, V883 Ori and HBC 722 and then comparing the predicted spectrum to observed optical and infrared spectra. 
The brightness ratio between the viscous disk and the stellar photosphere, $\eta$, provides an important guide
for identifying viscous accretion disks, with 
$\eta=1$ (``transition line") and $\eta=5$ (``sufficient dominance line") marking turning points in diagnostics, 
evaluated here in the near-infrared.
These turning points indicate the emergence and complete development of FUor-characteristic strong CO absorption, weak metallic absorption, the triangular spectral continuum shape in the $H$-band, and location in color-magnitude diagrams.  Lower 
$M_*$ and higher 
$\dot{M}$ imply larger $\eta$; for $M_*=0.3~{\rm~M_\odot}$, $\eta=1$ corresponds to $\dot{M}=2\times10^{-7}~{\rm~M_\odot}/$yr and $\eta=5$ to $\dot{M}=6\times10^{-7}~{\rm~M_\odot}/$yr. The sufficient dominance line also coincides with the expected accretion rate where
accreting material directly reaches the star.  We discuss implications of the models on extinction diagnostics, FUor brightening timescales, viscous disks during initial protostellar growth, and eruptive young stellar objects (YSO) associated with FUors.
	
\end{abstract}


\section{Introduction} \label{intro}


The role of outbursts in the mass assembly of a star 
remains uncertain \citep[see reviews by][]{dunham14,fischer22}, despite important implications for the evolution of the protoplanetary disk, envelope, and the protostar itself.  The largest eruptions during star formation, FUor-type outbursts (FUors), exhibit bursts of $\sim 5$ mag at optical wavelengths that last for decades \citep[see reviews by][]{hartmann96,audard14}.  Most of the handful of bona fide FUors have been discovered in optical monitoring surveys, with very low occurrence rates \citep{contreras19}.  However, young protostars are still embedded in envelopes and are not optically visible.  Infrared and sub-mm monitoring experiments developed in the past decade have suggested that outburst rates may be higher for this younger set of stars \citep[e.g.,][]{fischer19,guo20,lee21,park21,zakri22}.


Since many FUors, including FU Ori itself, decay on century timescales, some young stellar objects (YSOs) should already be in outburst, even though the initial rise was not detected.  Most bona fide FUor outbursts have
luminosities greater than 50 L$_\odot$  \citep[e.g.,][]{connelley18}, leading to accretion rates sufficient to viscously heat the disk at small radii to 5000--10000 K.
Ongoing outbursts can therefore be discovered by searching for the distinct spectral features of the heated inner disk of FUors; 
these objects are called {\it FUor-like} objects, with a putative burst that would have occurred prior to any monitoring \citep[e.g.][]{beichman81,reipurth02,aspin03,greene08}.
The category of FUor-like disks consists of tens of objects with unmistakable spectral features from a hot disk, including a few mysterious objects with bolometric luminosities $<10$~L$_\odot$ (many labeled  ``peculiar" by \citealt{connelley18}; see also, e.g., \citealt{yoon21}).  Such a faint luminosity indicates a low accretion rate $\sim 10^{-7}$~M$_\odot$ yr$^{-1}$, consistent with the upper envelope of Class \uppercase\expandafter{\romannumeral2} disks \citep{manara17},  objects with disk heating dominated by stellar radiation.



These previous studies motivate us to evaluate various contributions to optical and near-IR spectra across a wide range of accretion rates and masses.  Our focus is the viscously heated inner disk, as initially developed by \citet{Calvet91}, and following a modernized spectroscopic treatment by \citet{zhu07} and \citet{rodriguez22}; similar approaches have been developed for photometry \citep{gramajo14,kospal16}. 
Our models also include magnetospheric accretion, the passively heated disk, and the stellar photosphere so that we can calculate the relative contributions of each component across a wide parameter space.
On the low end of accretion rates, namely for T Tauri stars (TTSs), the stellar photosphere dominates spectral features, along with the magnetospheric accretion and dusty circumstellar disk irradiation that respectively explain the UV/optical and infrared excess emission \citep{hartmann94,chiang1997}. These components are incorporated in pre-main sequence (PMS) evolutionary contexts.  
In Sect.~\ref{models of emission components}, we assemble spectra with models of emission components as a function of stellar mass and accretion rate. 
In Sect.~\ref{parameter space of the disk model}, we discuss the parameter space for the models, before validating the models in Sect.~\ref{validation of model}. In Sect.~\ref{observational diagnostics}, we analyze observational characteristics of FUors through the parameter space. In Sect.~\ref{discussion}, we discuss implications for detecting FUors and the possibility of viscous disks early in the formation of a protostar.  
We summarize our results in Sect.~\ref{conclusions}.

\section{Models of emission components}
\label{models of emission components}

\begin{figure*}
    \centering
    \includegraphics[width=0.8\textwidth]{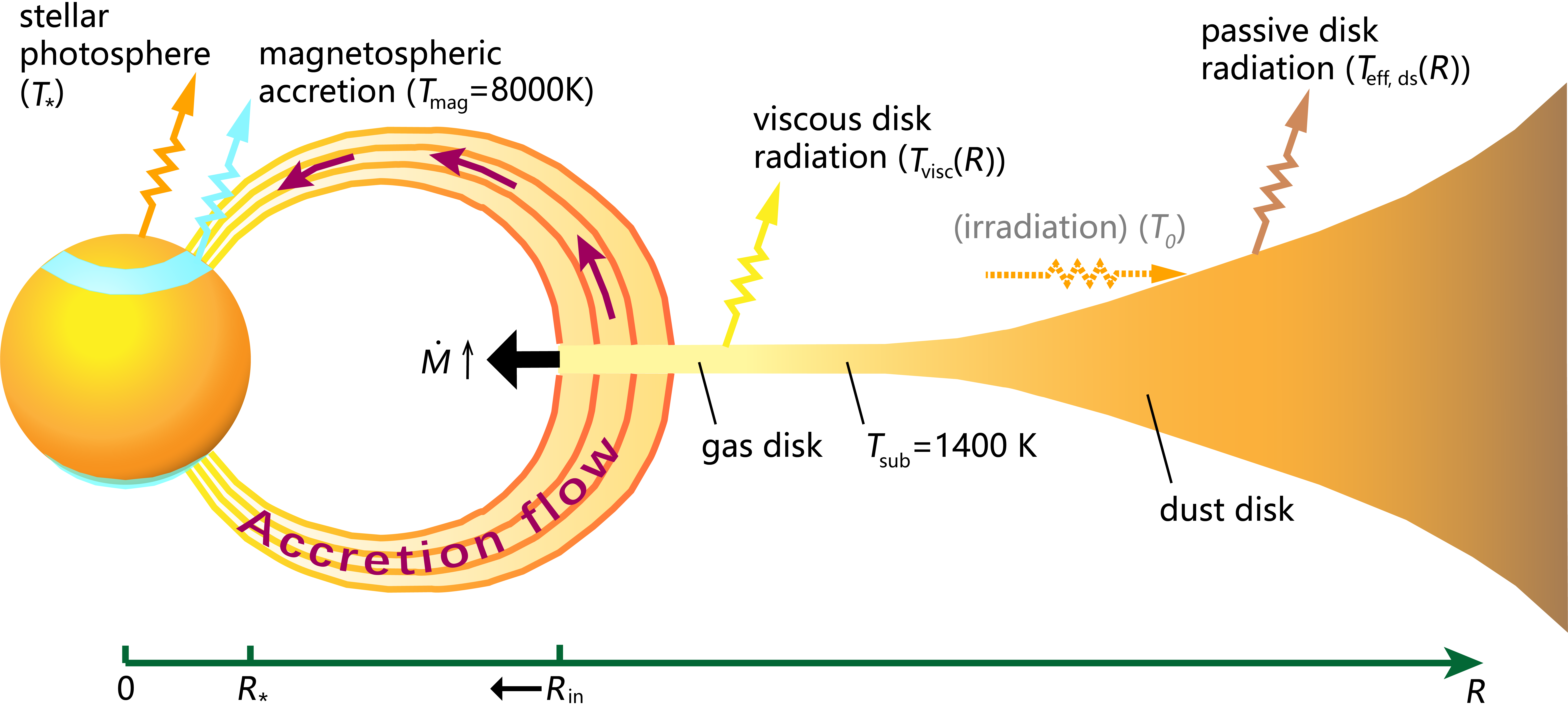}
    \caption{{Schematic view of the star-disk system, adapted from Fig.~1 in \cite{hartmann16}, with magnetospheric accretion and the four emission components (shown as squiggle arrows) in our models. The disk temperature decreases with the radius, forming an inner gas disk where the temperature is hotter than the dust sublimation temperature $T_{\rm sub}=1400$ K and a dust disk further outside. The gas disk is assumed to be geometrically thin. If the accretion rate $\dot{M}$ increases, the disk inner boundary will move inward until disk material directly reaches the star (see definition of $R_{\rm in}$ in Sect.~\ref{subsec viscous disk}).}}
    \label{fig model_schematic}
\end{figure*}

In this section, we describe the primary components of optical and infrared emission from young stars: the photosphere of viscous FUor disks, magnetospheric accretion, the stellar photosphere, and warm dust. We illustrate our general setup in Fig.~\ref{fig model_schematic}. The setup is idealized, with simplified geometries to approximate complex structures.  For example, the analytical formula for a viscously heated disk assumes that the disk is geometrically thin, even though the innermost disk is likely thick.  The star-disk interaction region is also complicated; the attenuation of X-rays from FUor objects may even indicate that the inner disk envelops the star \citep{kuhn19}.   We do not consider wind absorption, although P Cygni absorption profiles are commonly detected in strong lines \citep[e.g.][]{calvet93,lee15,milliner19}. We also do not include molecular emission from the irradiated disk, though CO overtone emission is important strong accretors, including EX Lup-type objects \citep{najita96,aspin10}, and could in principle fill in CO absorption for FUors.
Despite these simplifications, also adopted by other treatments \citep[e.g.][]{zhu07,rodriguez22}, our calculations provide guidance for the parameter space when a viscously heated disk matters in the optical and near-IR emission.

\subsection{FUor disk photospheres}
\label{subsec viscous disk}
We employ the standard $\alpha$-disk model, following \cite{shakura73}. We assume a steady, geometrically thin, and optically thick accretion disk, with the disk material orbiting the central star in pure Keplerian motion. The disk is heated by the viscous force due to the difference in angular velocity between different annuli. The effective temperature distribution in the disk is a function of the radial distance $R$, given by
\begin{equation}
    T_{\rm visc}^4(R)=\frac{3GM_*\dot{M}}{8\pi \sigma R^3}\left[1-\left(\frac{R_{\rm in}}{R}\right)^{1/2}\right]
    \label{eq Tviscous}
\end{equation}
where $M_*$ is the stellar mass, $\dot{M}$ is the accretion rate, $\sigma$ is the Stefan-Boltzmann constant, and \ri \ is the inner boundary radius of the disk.

The disk extends inwards to either the stellar photosphere or an inner truncation radius.  In the case of mild accretion, the disk is truncated by the magnetosphere, where the disk ram pressure is balanced by the magnetic field pressure:
\begin{equation}
    R_{\rm in} = \frac{B_*^{4/7}R_*^{12/7}}{\dot{M}^{2/7}(2GM_*)^{1/7}}
    \label{eq truncation}
\end{equation}
where $B_*$ is the stellar field strength, assumed as dipole, and $R_*$ is the stellar radius  \citep[see reviews by][]{bouvier07,hartmann16}. For simplicity, we fix $B_* = 1$\ kG, in {approximate} agreement with Zeeman broadening and Zeeman Doppler Imaging of classical TTSs  \citep[e.g.,][]{johnskrull07,donati09}. In the case of strong accretion, Eq.~\ref{eq truncation} yields a disk radius smaller than $R_*$. The corresponding physical image is that the gas pressure overwhelms the magnetospheric pressure, so that the disk material directly hits the star. In this case, we thus set \ri = $R_*$. Overall,
\begin{equation}
    R_{\rm in} = \max\left\{\frac{B_*^{4/7}R_*^{12/7}}{\dot{M}^{2/7}(2GM_*)^{1/7}}, R_*\right\}
    \label{eq rin}
\end{equation}
as in Eq.~\ref{eq truncation}.

A cautionary note is that in Eq.~\ref{eq Tviscous}, the temperature will unphysically drop to 0 K when $R$ approaches \ri\ \citep[the boundary layer problem for FUors;][]{kenyon88,popham96}. 
 For simplicity, we assume that within 1.36~\ri , the radius where $T_{\rm visc}(R) = T_{\rm max}$ reaches the maximal value, the temperature remains constant at \tmax\ \citep[following similar approaches by, e.g.,][]{kenyon88,zhu07}. {For a given protostar with $M_*$ and $R_*$ held constant, \tmax\ positively depends on the accretion rate, as
 \begin{equation}
    T_{\rm max}=6840~{\rm K} \left(\frac{M_*}{0.3~{\rm M}_\odot}\frac{\dot{M}}{10^{-5}~{\rm M}_\odot/ {\rm yr}} \left(\frac{1.6~{\rm R}_\odot}{R_{\rm in}}\right)^3\right)^{1/4}
\end{equation}
where \ri\ diminishes or remains invariant with the increase of $\dot{M}$.} 
 FUor disks have \tmax\ of several to roughly 10,000 K (see Table \ref{tab accretion reference} and references therein for examples).

We denote the sublimation temperature of the circumstellar dust as $T_{\rm sub} = 1400$~K {\citep{posch07}}. For the gaseous regions with temperature $T_{\rm visc} > 1400$~K, we assume that the emission resembles a stellar photosphere of a low gravity giant or supergiant, following empirical descriptions \citep[see, e.g.,][]{hartmann96,rodriguez22}.  We use the BT-Settl model grid of theoretical spectra (using solar AGSS2009 abundances), taking advantage of the fine scale in temperatures \citep{allard12}.
For gravity, at $1400~{\rm K} < T_{\rm visc} < 2000~{\rm K}$ we adopt 
$\log g = 3.5$, the lowest gravity templates available, and at $T_{\rm visc} \geq 2000$~K we adopt $\log g = 1.5$.  
We confirm that these models broadly reproduce observed high-resolution infrared spectra of giant stars obtained by \citet{park18}.  For regions with temperature $T_{\rm visc} \leq 1400$ K, we assume that the warm dust dominates the radiation and thus obscures the viscous emission (the warm dust will be treated in Section \ref{subsec warm dust}). This truncation formally sets an outer boundary radius, $R_{\rm sub}$, for the disk photosphere model at $T_{\rm visc}(R_{\rm sub}) = T_{\rm sub}$ = 1400~K.
We then divide the accretion disk into narrow annuli with radius $R$ and width $dR$, and calculate the contribution of each annulus from the annular temperature, rounded to the nearest available template, and from the surface area of the emission.


A characteristic feature of the disk emission spectrum is the broadened line profile from the fast Keplerian rotation. To reproduce this effect, we convolve each annular spectra with
\begin{equation}
    \phi (\lambda) = \left[1 - \left(\frac{\lambda - \lambda_0}{\lambda_{\rm max}}\right)^2\right]^{-1/2}
\label{eq broadening}
\end{equation}
to apportion the specific luminosity at $\lambda_0$ in a wavelength interval, with
\begin{equation}
    \lambda_{\rm max} = \lambda_0 \frac{\sin i}{c}v_{\rm Kep} = \lambda_0 \frac{\sin i}{c}\sqrt{\frac{GM_*}{R}}
\end{equation}
where $i$ is the inclination angle of the disk.

After broadening the spectra for disk rotation, we sum the contribution from each annulus as a function of wavelength. The specific photospheric luminosity $L_\lambda$ of the entire disk (on one side) is given by
\begin{equation}
   L_\lambda d\lambda  = \int_{R_{\rm in}}^{R_{\rm sub}}F_{\lambda}(\lambda,T_{\rm visc})2\pi R d R d\lambda .
\end{equation}
Denoting the distance of the system as $d_*$, we finally obtain the observed spectral flux
\begin{equation}
F_{{\rm obs}, \lambda}d\lambda = \frac{L_\lambda d\lambda \cos{i}}{\pi d_*^2}.
\label{eq viscous flux}
\end{equation}

\subsection{Magnetospheric accretion}
\label{subsec mag accretion}
The magnetospheric accretion model of T Tauri disks \citep[e.g.,][]{konigl91,hartmann94} describes that disk material freely falls onto the stellar surface, releasing kinetic energy.  For simplification purposes, we assume that the {energy} escapes in the form of blackbody radiation with an effective temperature of $\sim 8000$~K, estimated from modeling excess line and continuum emission \citep[see review by][]{hartmann16}.
We apply this framework to our accretion disk model for parameters where we expect magnetopsheric accretion, with the magnetospheric accretion luminosity $L_{\rm mag}$ of
\begin{equation}
    L_{\rm mag} = \frac{G\dot{M}M_*}{R_*}\left(1-\frac{R_*}{\langle R_{\rm M}\rangle}\right).
    \label{eq mag}
\end{equation}
where $\langle R_{\rm M}\rangle$ denotes the average radius on the disk where the accretion flow starts. We approximate $\langle R_{\rm M}\rangle$ with $R_{\rm in}$, with errors to be discussed later in this subsection. The luminosity implies an accretion shock area
\begin{equation}
    A_{\rm shock} = \frac{L_{\rm mag}}{\sigma T_{\rm mag}^4}
\end{equation}
where $T_{\rm mag} = 8000$~K. Eq.~\ref{eq mag} also implies nonexistence of the magnetospheric accretion emission if $R_{\rm in} = R_*$.

\begin{figure}
    \centering
    \includegraphics[scale=0.13]{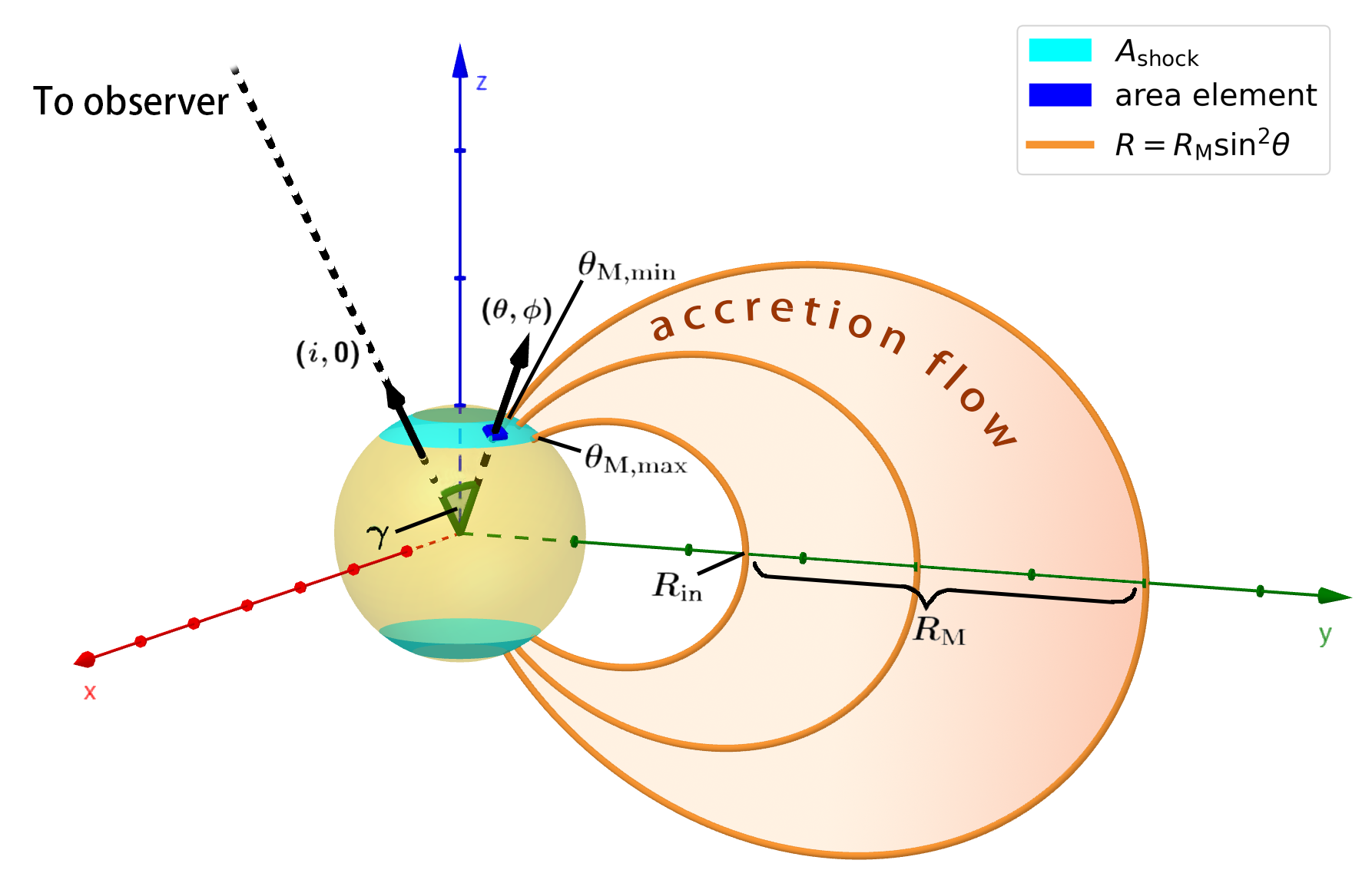}
    \caption{{Schematic view of the magnetospheric accretion. The accretion flow, displayed as a slice in the yz-plane, starts from the disk surface (not shown, in the xy-plane) at a range of radii ($R_{\rm M}$, curly bracket), follows trajectories specified in Eq.~\ref{eq trajectory},  and reaches the star at a range of polar angles ($\theta_{\rm M}$, with $\theta_{\rm M,min}$ and $\theta_{\rm M,max}$ annotated). The accretion shock area $A_{\rm shock}$ is colored in cyan on the stellar surface. In Eq.~\ref{eq magnetospheric flux}, the flux is integrated from area elements (blue) with the normal vector $(\theta, \phi)$. The normal vector forms an angle $\gamma$ with the normalized vector pointing to the observer, which has spherical coordinates $(i,0)$.}}
    \label{fig magnetosphere schematic}
\end{figure}

We further assume a dipole stellar magnetic field in line with the stellar rotational axis (e.g., \citealt{alencar12}; see also discrepancies described by \citealt{adams12}). {The accretion flow starts from the disk surface at a range of radii. Let $R_{\rm M}$ be a specific radius in this range, and the accreting material follows} the trajectory in spherical coordinates
\begin{equation}
    R = R_{\rm M}\sin^2\theta
    \label{eq trajectory}
\end{equation}
and lands on the star at {an $R_{\rm M}$-dependent} polar angle
\begin{equation}
    \theta_{\rm M}(R_{\rm M}) = \arcsin \sqrt{R_*/R_{\rm M}}.
\end{equation}
The disk inner radius $R_{\rm in}$ specifies the minimal value of $R_{\rm M}$, which when combined with $A_{\rm shock}$ allows us to infer the {$\theta_{\rm M}$ range and the $R_{\rm M}$ range} of the accretion shock. Denoting the minimal and maximal shock polar angle on the northern hemisphere as {$\theta_{\rm M, min}$ and $\theta_{\rm M, max}$}, we calculate the observed flux
\begin{equation}
\begin{split}
    &F_{{\rm obs}, \lambda}d\lambda = \left(R_*^2/d_*^2\right)\cdot B(\lambda, T_{\rm mag})d\lambda\\
    \cdot&\int_0^{2\pi}d\phi \left(\int_{\theta_{\rm M, min}}^{\theta_{\rm M, max}} + \int_{\pi-\theta_{\rm M, max}}^{\pi-\theta_{\rm M, min}}\right)\cos\gamma\sin\theta d\theta
\end{split}
\label{eq magnetospheric flux}
\end{equation}
where $B(\lambda, T_{\rm mag})d\lambda$ is the blackbody spectral radiance at $T_{\rm mag} = 8000$ K, and 
\begin{equation}
\cos\gamma = \max \left\{\sin\theta\cos\phi{\sin i} + \cos\theta\cos i,\ 0\right\}
\end{equation}
{is the inner product of the normalized vector towards the observer and the normal vector of an area element at spherical coordinates $(\theta, \phi)$, where $\theta$ is the polar angle and $\phi$ the azimuthal angle from the x-axis. Only area elements with positive $\cos\gamma$ are not blocked by the star itself. The geometrical settings are shown in Fig.~\ref{fig magnetosphere schematic}.}

For typical TTS parameters of $M_* = 0.3$~M$_\odot$, $R_* = 1.6$~R$_\odot$ and $\dot{M} = 1\times10^{-8}$~M$_\odot$/yr, $\theta$ spans from $48^\circ$ to $50^\circ$, covering 3\% of the stellar surface, roughly consistent with estimated hot spot fractions in previous studies \citep[e.g.][]{calvet98}. However, the real distribution of these hot spots may deviate from the simplistic picture established here: the existence of higher-order magnetic fields and the misalignment between the dipole and the stellar rotational axis both lead to distribution of spots of heterogeneous polar and azimuthal angle
dependence.  We ignore alternative geometries and formulations, including equatorial flows at high accretion rates \citep{romanova08} and flows that are not magnetospheric because of weak magnetic fields \citep{takasao18}.
Our simplistic approach is intended to provide guidance rather than detailed accuracy.



\subsection{Stellar photosphere}
\label{stellar photosphere}

We assume that the photosphere is well approximated with a BT-Settl template spectrum of a single effective surface temperature $T_*$ over the stellar surface area. This treatment (ignoring the disturbance of the photosphere by direct or magnetospheric accretion) is justified by the small fraction of surface area covered by accretion columns \citep[e.g.][]{ingleby13,robinson19}.  We fix the photospheric gravity $\log g = 3.5$, which agrees with standard PMS models \citep{baraffe15} and roughly with spectroscopic measurements \citep[e.g.,][]{cottaar14}. We ignore rotational broadening of the photospheric spectra, as line features are dominated by the viscous-disk emission in our scope of interest.

The detectability of the accretion disk depends on the PMS evolution of the central star. We place the disk model into the evolutionary context both as a way to add reasonable constraints to the parameter space and as a lens to better understand the significance of FU Ori bursts to stellar growth. For low-mass stars, we employ a standard, non-magnetic model from \cite{baraffe15}. Our fiducial models for this paper adopt $T_*$ and $R_*$ 
for a range of stellar masses at an age of 1~Myr.


We also extend the upper boundary of the stellar mass to explore the behavior of young intermediate-mass stars interacting with circumstellar disks. To stay consistent with the standard \cite{baraffe15} model, we employ the non-magnetic 1~Myr isochrone from \cite{feiden16} to cover the mass range from 1.5--3.0~M$_\odot$.

{Fully convective stars are often heavily spotted.  Our evaluations of the spectroscopic diagnostics at high accretion rates are robust to differences in the underlying photospheric spectrum.  Even for extreme cases where the cool spot covers $\sim80\%$ of the stellar surface \citep[e.g.,][]{gully-santiago17}, two-temperature models does not differ significantly from a single average temperature model in terms of SED and line depth in the near-IR, except at low accretion rates.}





\subsection{Warm dust in the disk}
\label{subsec warm dust}

\begin{figure*}[!th]
    \centering
    \includegraphics[scale=0.55]{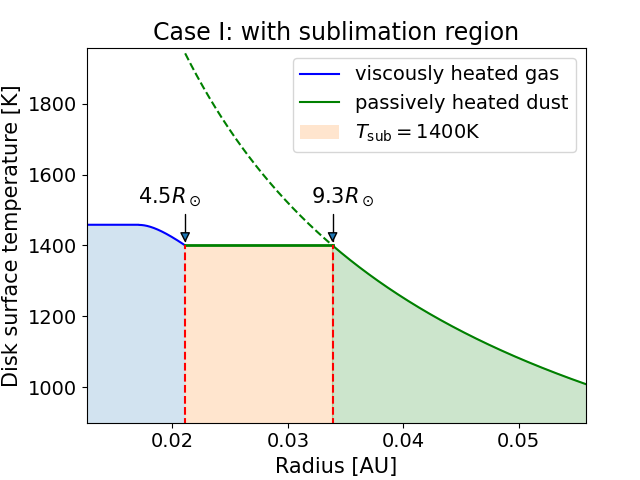}
    \includegraphics[scale=0.55]{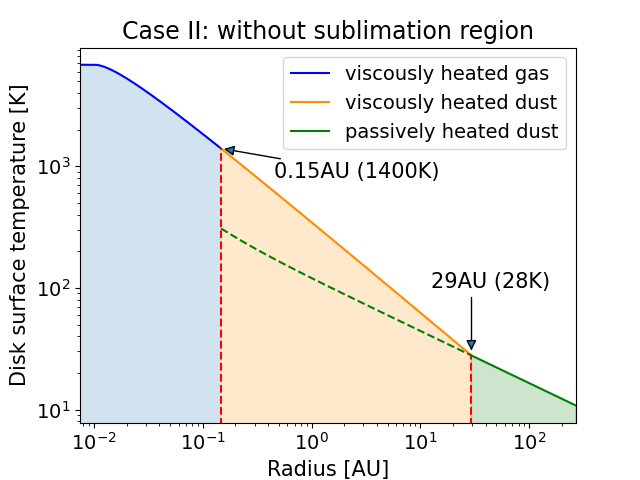}
    \caption{{Disk surface} temperature versus radius with a sublimation region for a two models of a $M_* = 0.3 \ {\rm M_\odot}$ star, on the left with  $\dot{M} = 1\times10^{-7} {\rm M_\odot} / {\rm yr}, \ R_* = 1.6 {\rm R_\odot}, R_{\rm in} = 2.7 {\rm R_\odot}, \ T_0 = 6000$~K and on the right with  $\dot{M} = 1\times10^{-5} {\rm M_\odot} / {\rm yr}, \ R_* = R_{\rm in} = 1.6 \ {\rm R_\odot}, \ T_0 = 3400$~K.
    The disk with the lower accretion rate ({\it left}) is cut into three rings: the inner gaseous region, with luminosity mainly from the viscous heating; the intermediate sublimation region, which radiates like a 1400~K blackbody; the outer dusty region, with luminosity mainly from the superheated surface dust layer.   The disk with the higher accretion rate ({\it right}) 
    is cut into three rings: the inner gaseous region, keeping its temperature via viscous heating; the hotter dusty region, keeping its temperature via viscous heating; the cooler dusty region, receiving energy from the stellar radiation.}
    \label{fig cases}
    \vspace{1em}
\end{figure*}

We follow the \cite{chiang1997} model for dust in the disk, at disk locations where the surface temperature is cooler than $\sim1400$~K. In this mode, the surface of the disk is heated by the stellar radiation, which then radiates like a blackbody. The effective temperature of the heated surface at radius $R$ obeys
\begin{equation}
    T_{\rm eff, ds}(R) = \left(\frac{\alpha_0}{2}\right)^{1/4} \left(\frac{R_*}{R}\right)^{1/2} T_0,
    \label{eq Tdust}
\end{equation}
where $\alpha_0$ is the angle between the incident stellar light and the disk surface and $T_0$ denotes the effective temperature of the radiation that heats the dusty surface rather than the temperature of the stellar photosphere.
The radiation may stem from the stellar photosphere, the magnetospheric accretion, and even the bright viscous disk. The geometrically thin assumption may collapse near the inner boundary, where the hot disk material possibly piles up, deviates from the surface plane, and radiates part of its energy towards the dust. In sum, the physics near the stellar surface is complicated, and we wrap everything up in the single parameter of $T_0$ to draw a simple picture.
The value of $T_0$ is at minimum the photospheric temperature $T_*$, and may be higher if there are additional contributions to the luminosity and disk heating.

For a disk in hydrostatic equilibrium, $\alpha_0$ depends on the radial location of the disk. In general, 
\begin{equation}
    \alpha_0 = \frac{0.4R_*}{R} + R\frac d{dR}\left(\frac{H}{R}\right)
\end{equation}
where $H$ is the 
disk height above the midplane. Under the assumption that the dust and gas are uniformly mixed in the outer disk, the angle $\alpha$ can be numerically expressed as
\begin{equation}
    \alpha_0 = 0.003\frac{R_{*1.6}}{R_{\rm AU}} + 0.05\frac{T_{*3400}^{4/7}R_{*1.6}^{2/7}}{M_{*0.3}^{4/7}}R_{\rm AU}^{2/7}
\end{equation}
where fiducial values for parameters $R_*, T_*$ and $M_*$ are noted in subscript, and $R_{\rm AU}$ is the disk radius measured in AU.

We set the outer boundary of the dust disk uniformly as 270 AU, following \cite{chiang1997}. Different outer radii do not significantly affect the near- or mid-IR emission.


Our analytical description of the disk temperature applies Eq.~\ref{eq Tviscous} to obtain a viscous temperature $T_{\rm visc}$ for the inner disk and Eq.~\ref{eq Tdust} to obtain a dust effective temperature $T_{\rm eff, ds}$ for the surface temperature of the outer disk.  The connection between these two regions is unclear. Allowing all free parameters to vary, we identify two cases, both based on arguments for the dust sublimation temperature $T_{\rm sub}$.


\textbf{Case \romanone}, where $T_{\rm visc}$ reaches $T_{\rm sub}$, $T_{\rm eff, ds} > T_{\rm sub}$, Fig.~\ref{fig cases} ({\it left}): in the illustrated example, $T_{\rm visc} < T_{\rm sub} = 1400~{\rm K} < T_{\rm eff, ds}$ between 4.5 ${\rm R_\odot}$ and 9.3 ${\rm R_\odot}$. We reason that a phase-transition state shows up in this region. If the surface temperature is higher than 1400~K, then the dust grains will sublimate, leaving behind the gas, which is optically thin at most infrared wavelengths where the warm dust emits.
The interior, heated by the viscous shear, is cooler than 1400~K and does not contribute heat to the surface. Based on these considerations, we assume that the temperature in the region is kept constant at $T_{\rm sub}$ and emits as a blackbody.

\textbf{Case \uppercase\expandafter{\romannumeral2}}, where $T_{\rm visc}$ reaches $T_{\rm sub}$, $T_{\rm eff, ds} < T_{\rm sub}$, Fig.~\ref{fig cases} ({\it right}): in the illustrated example, $T_{\rm eff, ds} < T_{\rm visc} < T_{\rm sub} = 1400$~K between 0.15 AU and 29 AU. This temperature is cool enough for dust to be present; the emission from this region is treated as optically thick blackbody emission from viscously heated warm dust.






\section{Parameter space of the disk model}
\label{parameter space of the disk model}
In this section, we explore the effects of varying the physical input parameters to evaluate spectral properties of the star-disk system. The assembly of parameters is shown in Table \ref{tab fiducial}, along with a set of fiducial values for YSOs with high accretion rates.

\begin{figure*}[!t]
    \centering
    \includegraphics[scale=0.27]{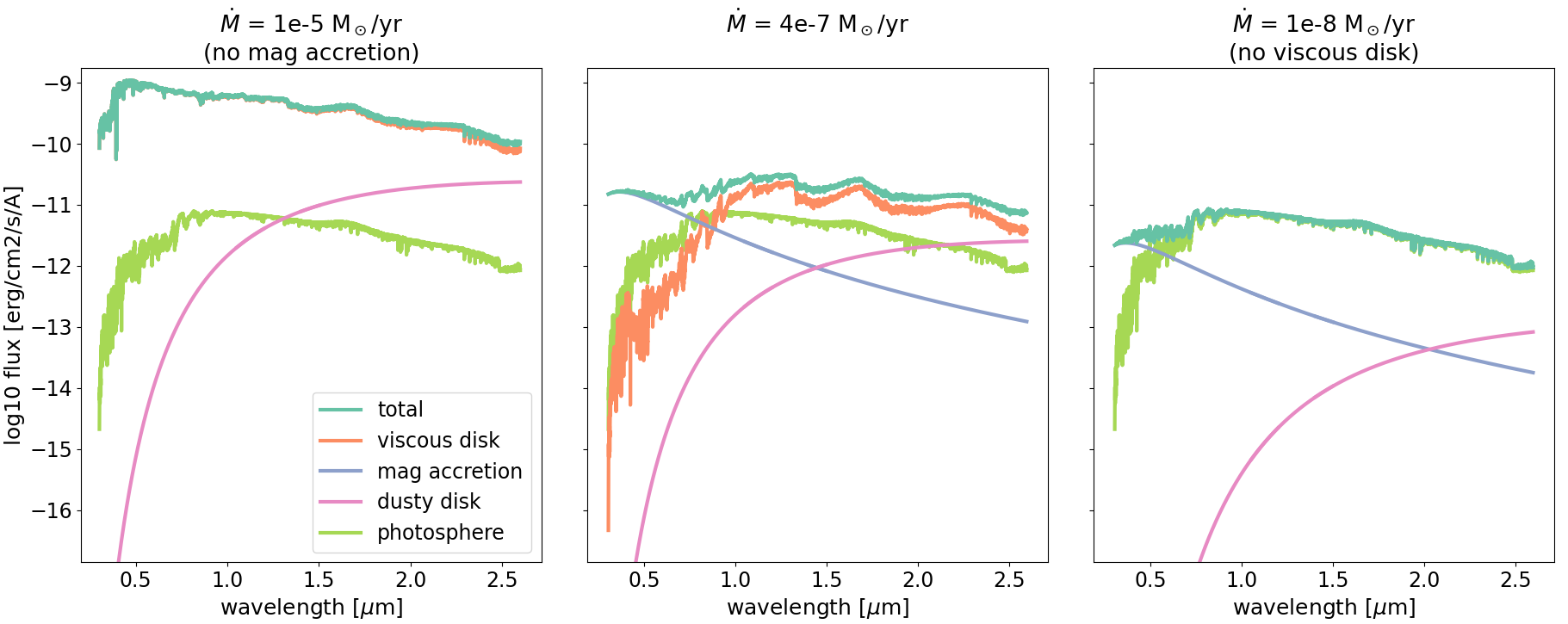}
    \caption{Components of $M_*=0.3~{\rm M_\odot}$ model spectra across different accretion rates. High accretion rates (e.g., $10^{-5}~{\rm M_\odot}/$yr) crush the magnetosphere and eliminate magnetospheric accretion, while low accretion rates (e.g., $10^{-8}~{\rm M_\odot}/$yr) fail to heat up the disk viscously. The dust disk component mainly contributes to the continuum in the near-IR. The magnetospheric accretion, when existent, affords optical excess to the SED. Line features vary from predominantly disk-sourced (high $\dot{M}$) to photosphere-sourced (low $\dot{M}$).}
    \label{fig components}
    \vspace{1em}
\end{figure*}

We display in Fig.~\ref{fig components} components of our model spectra scaled to a distance of 10 pc. The shape of our model spectra is mainly determined by three relations: the temperature distribution functions, Eq.~\ref{eq Tviscous} and Eq.~\ref{eq Tdust}, as well as the constraint from the PMS model. We proceed to investigate how variations of $M_*$, $\dot{M}$, $i$, $T_0$ and $A_V$ influence the SED and the apparent spectral type.

\begin{deluxetable}{cccccccc}[!b]
\label{tab fiducial}
\tablecaption{Parameter space of a fiducial spectrum (Case II)}
\tablehead{\colhead{$M_*$} & \colhead{$\dot{M}$} & \colhead{$i$} & \colhead{$T_0$} & \colhead{$A_V$} & \colhead{$T_*$\tablenotemark{$\dag$}} & \colhead{$R_*$\tablenotemark{$\dag$}} & \colhead{$R_{\rm in}$\tablenotemark{$\ddag$}} \\ 
\colhead{(${\rm M_\odot}$)} & \colhead{(${\rm M_\odot}$/yr)} & \colhead{} & \colhead{(K)} & \colhead{(mag)} & \colhead{(K)} & \colhead{(${\rm R_\odot}$)} & \colhead{(${\rm R_\odot}$)} } 

\startdata
0.3 & $1\times10^{-5}$ & $45^{\circ}$ & 3400 & 10.0 & 3400 & 1.6 & 1.6\\
\enddata
\tablenotetext{\dag}{Parameters constrained by the PMS model.}
\tablenotetext{\ddag}{Parameter constrained by Eq.~\ref{eq rin}.}
\vspace{-2.5em}
\end{deluxetable}


\subsection{Stellar mass \texorpdfstring{$(M_*)$}{}}
\label{3.1}
The central mass $M_*$ is the pivotal parameter, as it leads to an $R_*$ and $T_*$ at a given age and shapes the width of the disk-formed spectral lines via Keplerian rotational broadening (Eq.~\ref{eq broadening}). The mass of the central object also vitally controls the disk temperature distribution as the term $M_*\dot{M}$ in Eq.~\ref{eq Tviscous}.

\begin{figure}[ht]
    \centering
    \includegraphics[scale=0.52]{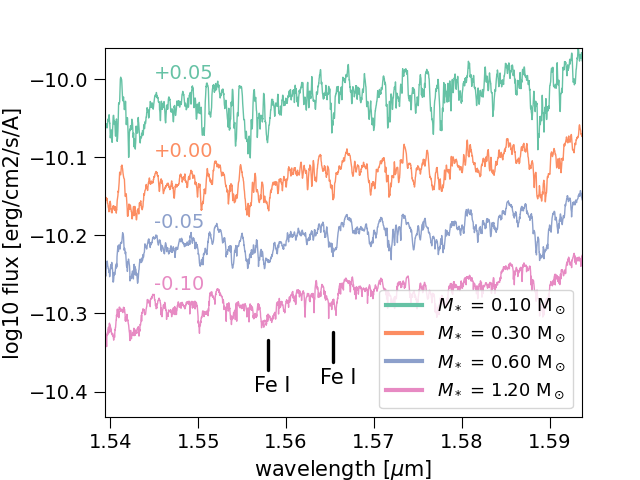}
    \caption{A small segment of model spectra of different stellar masses, with the product $M_* \dot{M} = 3.0\times10^{-6} {\rm M_\odot}^2$/yr, $i = 45^\circ$, $A_V = 10$ mag all held constant and $T_0$ set as the minimal value $T_*$ given by the PMS model.
    {The spectra are vertically shifted. They do not perfectly overlap before the offset because $T_{\rm visc}$ also depends on $R_{\rm in}$ in Eq.~\ref{eq Tviscous}.} The two \ion{Fe}{1} lines show that higher $M_*$ entails broader, shallower profiles. Some thin needle-shaped lines are present in the $M_* = 1.20 {\rm M_\odot}$ spectrum; they originate from higher photospheric luminosity predicted by the standard PMS model.}
    \label{fig para_space_m}
\end{figure}

The spectral result of various central masses is shown in Fig.~\ref{fig para_space_m}, holding $M_* \dot{M}$ constant. Metal absorption lines (e.g., 
\ion{Fe}{1}
lines at $\sim1.56~\mu$m) of more massive objects are broader and shallower due to faster Keplerian rotation, showing boxlike or double-edged profiles. Such line shapes are interpreted as evidence to distinguish FUors from other PMS stars \citep{hartmann96,lee15}.

\subsection{Accretion rate \texorpdfstring{$(\dot{M})$}{}}
\label{accretion rate}
The mass accretion rate ($\dot{M}$) and stellar mass ($M_*$) constitute the numerator of Eq.~\ref{eq Tviscous}. If we fix $M_*$ and $R_*$ (or age), then the temperature distribution of the viscous disk is mainly dependent on $\dot{M}$, or equivalently $M_* \dot{M}$; the $(R_{\rm in}/R)^{1/2}$ term at most weakly correlates with $\dot{M}$. In this case, then, the product $M_*\dot{M}$ controls the viscous disk SED and the spectral type. 

\begin{figure*}
    \centering
    \includegraphics[width=\textwidth]{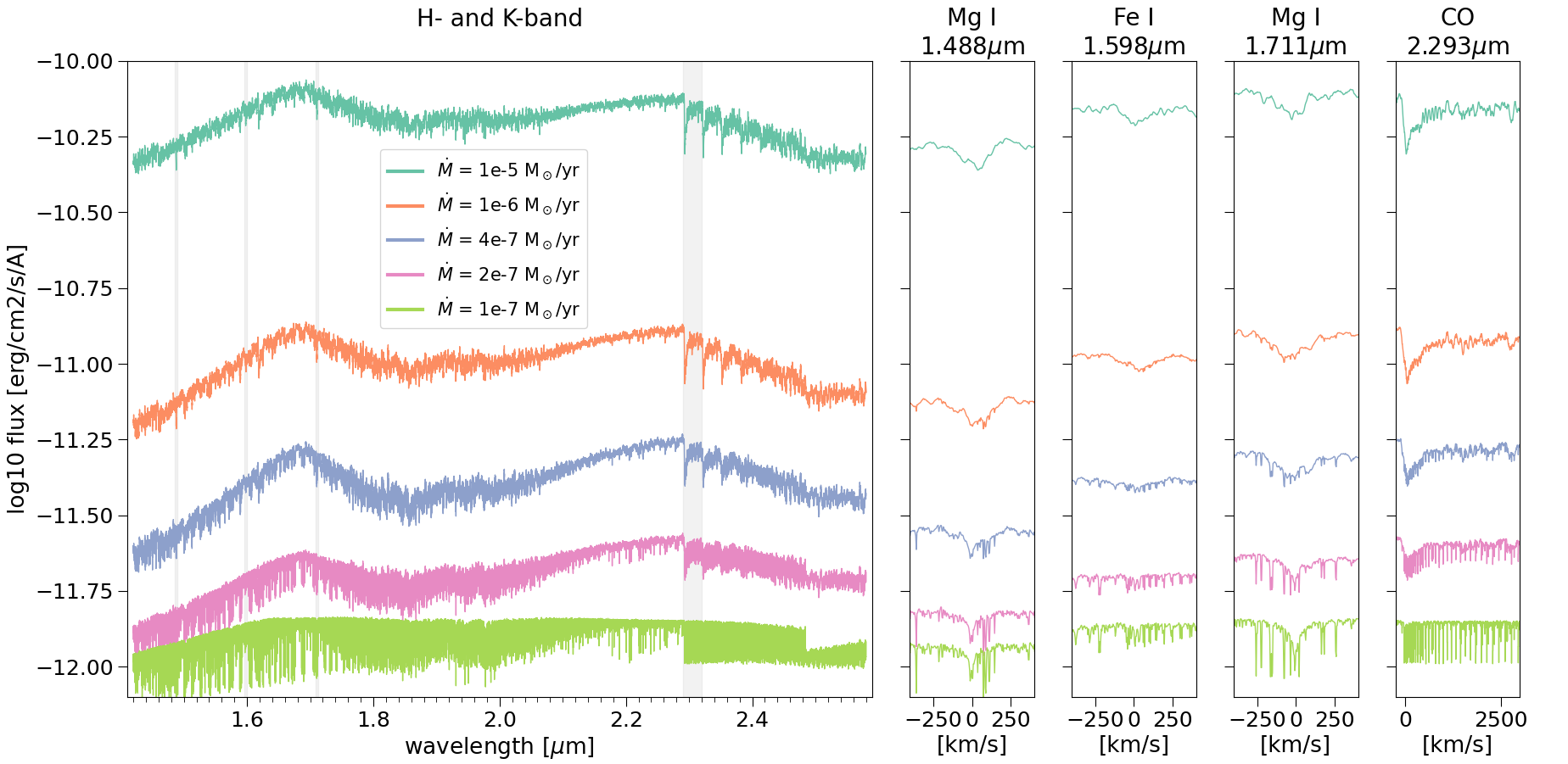}
    \caption{Model spectra of different accretion rates. Spectral fluxes are not shifted. The four shaded regions in the leftmost panel correspond to absorption lines in the right panels. While the total luminosity substantially changes with $\dot{M}$, specific line features such as Mg I, Fe I and CO lines are nearly unchanged where $\dot{M}\gtrsim10^{-6}\ {\rm M_\odot}/$yr. Lower accretion rates show gradual transition from disk to photospheric line features. Other parameters are held constant, with $M_* = 0.3~{\rm M_\odot}$, \ $i = 45^\circ$, $T_0 = 3400$~K, and $A_V = 10$~mag.}
    \label{fig para_space_mdot}
    \vspace{1em}
\end{figure*}

The model spectra for different accretion rates are shown in Fig.~\ref{fig para_space_mdot}. The total luminosity increases with higher $\dot{M}$, particularly as the maximum temperature of the viscous disk increases, which also leads to a larger surface area for emission from cooler temperature regions in the disk.
However, provided that $M_* \dot{M}$ is sufficiently high (e.g., larger than $3.0 \times 10^{-7}~{\rm M_\odot}^2$/yr), and that $M_*$ itself is not too large (so that the viscous disk becomes the predominant source of absorption lines), the spectral type from the near-IR spectrum will not vary significantly. The underlying reason is that a relatively high $M_* \dot{M}$ entails high \tmax \ for the viscous disk, which makes a wide range of temperature available for the viscous disk (see the blue region in Fig.~\ref{fig cases}, right panel). Therefore, spectral features of different temperatures may present themselves together. Moreover, the quasi-power-law temperature profile implies that the ratio of any two spectral type components with given temperatures, $(T_1^4R_1dR) / (T_2^4R_2dR)$, where $R_1\propto (M_*\dot{M}/T_1^4)^{1/3}$ and $R_2\propto (M_*\dot{M}/T_1^4)^{1/3}$  as in Eq.~\ref{eq Tviscous}, is independent from $M_*\dot{M}$.

In the near-IR region, it is mainly the less-altered lower temperatures that produce important features like metal and CO overtone absorption. Therefore, the lines in the spectra stay largely invariant when the accretion rate changes.
When conditions do not meet the criteria in the previous paragraph, the spectral line shapes will change because
(a) the stellar photospheric absorption lines begin to mix with the viscous disk lines, and (b) \tmax \ of the viscous disk is so low that even the modest temperature interval is disturbed.

\subsection{Stellar radius and the disk truncation radius}
We use the pre-main sequence models to link $R_*$ to $M_*$.  The inner disk boundary \ri\ is either directly determined by stellar radius $R_*$ in the case of strong accretion, or strongly correlates with $R_*$ in the case of mild accretion (Eq.~\ref{eq rin}). The close association between the two radii motivates us to discuss them in one subsection. The radius is a versatile and subtle factor: it influences the high end of the viscous-disk temperature and impact the luminosity of the magnetospheric accretion and the stellar photospheric radiation. 
The radius also affects the temperature distribution of the dusty disk surface. That the radius plays so many elusive roles in different aspects of the models discourages us from setting it as a free parameter. Hence, purporting to constrain $R_*$ and $R_{\rm in}$, we make the somewhat arbitrary assumption of 1~Myr stellar age in the PMS model. 
Apart from the age uncertainty, the application of the PMS model here ignores possible distension of the star in response to the high rate of material and energy inflow \citep{hartmann85}. We will return to the issue in Sect.~\ref{fits to data} when discussing fits to FUor photometry and spectroscopy

\subsection{Inclination angle \texorpdfstring{$(i)$}{}}
\label{inclination angle}
The system geometry, including the inclination angle, affects the observed flux from both the viscously heated and the irradiated disk. The observed width of  absorption line profiles from the disk also hinges on $i$. As for the stellar photospheric emission, on the other hand, the general observed flux is independent from $i$; still, the photospheric line widths vary with the angle.

The widths of absorption lines are vital criteria to assess goodness of fit; nevertheless, the challenge is that the inclination and the stellar mass constitute a {degenerate} relationship in terms of the line width (see Eq.~\ref{eq broadening}). 
Constraining $M_*$ from its subtle influence on other spectral features (discussed in Sect.~\ref{3.1}) is possible but
subject to large uncertainties; on the other hand, knowledge of the inclination from additional evidence will be of direct help to the fitting by improving its robustness.  {When available,} millimeter and near-IR interferometric images \citep[e.g., see][]{cieza16, labdon20} inform us of well-constrained inclination angles, which we conveniently adopt in our modeling. In the case of adopting ALMA results, we assume that the inner disk is in line with the outer disk. {When direct measurements of inclination are not available, the SED may provide loose constraints on the inclination.}


\subsection{Effective temperature for passive heating \texorpdfstring{$(T_0)$}{}}
\label{tzero}
Physically, radiative heating of the dusty disk consists of emission from the stellar photosphere, the magnetospheric accretion shocks, and presumably the viscous disk. The parameter $T_0$ includes each of these components, with a relative importance that depends on the emission strength.
The geometric structure of the innermost disk is particularly unsettled: it is still uncertain how large the scale height of this zone is compared with the stellar radius, and how its radiation is distributed to different directions. Therefore, we only constrain $T_0$ no more strictly than a very rough range, between the minimal and the maximal value among $T_*$, $T_{\rm mag} = 8000$ K, and \tmax.

\begin{figure}[t]
    \centering
    \includegraphics[scale=0.52]{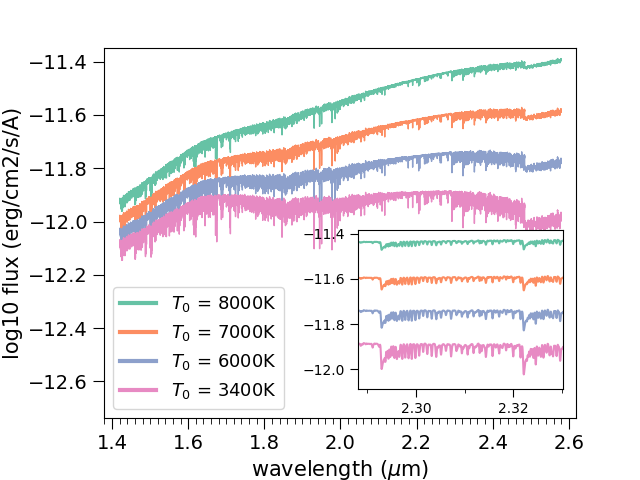}
    \caption{Model spectra of different $T_0$. For visualization, spectra are convolved to lower resolution. Higher $T_0$ helps to raise the SED, especially in the redder zones in the $K$-band, and veils the CO bandhead lines (enlarged on the lower right). Other parameters are held constant at $M_* = 0.3 {\rm M_\odot}$, $\dot{M} = 1\times10^{-7} {\rm M_\odot}/$yr, $i = 45^\circ$, $A_V = 10$ mag.}
    \label{fig para_space_T0}
\end{figure}

In Fig.~\ref{fig para_space_T0}, the hotter $T_0$ helps to raise the SED, especially in the redder zones in the $K$-band. This blackbody dust emission veils lines in the infrared, notably resulting in weakened CO bandhead lines for high $T_0$ values. However, $T_0$ remains influential only for moderate accretion rates or long wavelengths.  Given a high $\dot{M}$, even the dust disk can be predominantly viscously heated, which limits effects brought by the passive heating until mid- to far-IR.

This setup assumes that the inner viscous disk does not shadow the disk at larger radii \citep[e.g.,][]{okuzumi22}.  Alternate geometries may shield the disk at larger radii, in which case the contribution from warm dust would be smaller.

\subsection{Extinction from the envelope and interstellar medium}
\label{extinction}
Many FUor objects are heavily extincted, both because many are young enough to retain dense envelopes and because all are located in dusty star-forming regions. We thus redden our optical and near-IR model spectra with the \citet{Fitzpatrick_2019} (F19) model from \texttt{dust\_extinction}, a Python package affiliated to \texttt{Astropy}. In the model, we fix the parameter $R_V = 3.1$ in general, in accordance with the mean Galactic extinction law, while varying $A_V$, the total extinction at $V$. We vary $R_V$ only for V883 Ori to improve the fit to photometry  (see Sect.~\ref{fits to data}). 
{Here, diversity of the dust grain size implies uncertainties in $R_V$ and thus in extinction curve shapes.} At mid-IR wavelengths ($\lambda > 3.3~\mu$m), where the F19 curve no longer applies, we use the \citet{Gordon21} Milky Way extinction curves (also from \texttt{dust\_extinction}), which is highly consistent with F19 from ultraviolet through near-IR.

A high $T_0$ tends to brighten the red tail of the SED; in contrast, a high $A_V$, while weakening the flux at all wavelengths, primarily drags down the blue end. Therefore, the two parameters of $T_0$ and $A_V$ play important roles in shaping the overall trend of the spectra. {Since 
$M_*\dot{M}$ dominates the viscous disk temperature, the $\dot{M}$ and $A_V$ are also degenerate parameters,  with faster accretion and weaker extinction both leading to a bluer spectrum.}

\begin{deluxetable}{cccc}[b]
\label{tab photometry}
\tablecaption{Photometry data sources used for validation}
\tablehead{\colhead{Band} & \colhead{$\lambda_{\rm ref}$ ($\mu$m)} & \colhead{Survey/Telescope} & \colhead{Reference}}
\startdata
$B$ & 0.437 & Konkoly & \citealt{kospal11}\\
$G_{\rm BP}$ & 0.511 & {\it Gaia} (EDR3) & \citealt{GAIA} \\
$V$ & 0.548 & Konkoly & \citealt{kospal11}\\
$R$ & 0.650 & Konkoly & \citealt{kospal11}\\
$G_{\rm RP}$ & 0.777 & {\it Gaia} (EDR3) & \citealt{GAIA} \\
$I$ & 0.802 & Konkoly & \citealt{kospal11}\\
$J$ & 1.24 & 2MASS & \citealt{2MASS} \\
 & 1.26 & TCS & \citealt{kospal11} \\
$H$ & 1.65 & 2MASS & \citealt{2MASS} \\
 & 1.66 & TCS & \citealt{kospal11} \\
$K_{\rm S}$ & 2.16 & 2MASS & \citealt{2MASS} \\
 & 2.18 & TCS & \citealt{kospal11} \\
$W_1$ & 3.37 & WISE; NEOWISE & \citealt{Cutri:2012yCat} \\
$W_2$ & 4.62 & WISE; NEOWISE & \citealt{Cutri:2012yCat} \\
$I_3$ & 5.70 & Spitzer & \citealt{Spitzer/IRAC}\\
\hline
\multicolumn{4}{l}{{Konkoly: Schmidt Telescope at Konkoly Observatory}}\\
\multicolumn{4}{l}{{TCS: Telescopio Carlos Sanchez at Teide Observatory}}\\
\enddata
\vspace{-3.5em}
\end{deluxetable}

\begin{deluxetable*}{cccccc}
\tablecaption{Spectroscopic Observations}
\label{tab spectroscopy}
\tablehead{\colhead{Object} & \colhead{Date} & \colhead{Telescope} & \colhead{Instrument} & \colhead{$\lambda$} & \colhead{$\lambda/\delta\lambda$}}
\startdata
FU Ori & 2014-11-20 & McDonald 2.7m & IGRINS & 1.4--2.5 $\mu$m & 45000\\
V883 Ori & 2017-04-14, 2017-04-16 & McDonald 2.7m & IGRINS & 1.4--2.5 $\mu$m & 45000 \\
HBC 722 & 2014-11-19, 2014-11-20, 2015-11-23 & McDonald 2.7m & IGRINS & 1.4--2.5 $\mu$m & 45000 \\
HBC 722 & 2016-11-18 & Lowell 4 m & IGRINS & 1.4--2.5 $\mu$m & 45000\\
HBC 722 & 2016-12-14 & Palomar & TripleSpec & 0.98--2.5 $\mu$m & 2700\\
HBC 722 & 2012-08-07 & WHT & ISIS & 0.34--1.0  $\mu$m  & 1000
\enddata
\end{deluxetable*}


\section{Validating the models with fits to known FUors}
\label{validation of model}


In this section, we validate the models developed in Sect.~\ref{models of emission components}--\ref{parameter space of the disk model} by comparing them to the spectrophotometry of three FUor objects, FU Ori, V883 Ori and HBC 722.    We first describe the spectra and photometry and then use our models to reproduce the observations.  

\subsection{Spectroscopy and Photometry}

We collect near-IR and optical photometry for FU Ori, V883 Ori, and HBC 722 to validate our models for viscous disks. The photometry used in this paper is described in Table~\ref{tab photometry}, and the spectroscopic log is listed in Table~\ref{tab spectroscopy}.  High-resolution infrared spectra of FU Ori, V883 Ori, and HBC 722 were obtained with Immersion Grating Infrared Spectrograph \citep[IGRINS;][]{park14}.  The spectrum of HBC 722 was originally published by \cite{lee15}, while the spectra of FU Ori and V883 Ori are presented here for the first time.  

The spectra were reduced following standard procedures using the IGRINS pipeline \citep{Lee17}.  Telluric corrections were obtained from spectra of A0V standard stars observed just after or before each science target and at a similar airmass.  Flat-field calibration images were obtained to flatten the spectra, to identify and mask bad pixels, and to define the two-dimensional aperture of each cross-dispersed spectrum.  

One dimensional spectra are extracted by a modified version of the optimal extraction algorithm of \cite{Horne86}. The wavelength solution is derived from OH emission lines obtained from a blank sky each night. The telluric absorption lines from the A0V spectrum are used to refine the wavelength solution at longer wavelengths in the $K$-band because of the scarcity of the OH lines. We remove atomic hydrogen lines in the A0V spectrum by dividing it by a Vega model spectrum modified to fit the line profile of the A0V spectrum.  The A0V spectrum is then scaled by the differences in air mass between the A0V and the target observations. Finally, the telluric absorption lines in the target spectrum are removed by dividing the target spectrum by the A0V spectrum. We calibrate the flux of FU Ori and V883 Ori by scaling the spectra using 2MASS photometry \citep{park18}.

Low resolution near-IR spectroscopic observations of HBC 722 were acquired on 2016 December 14 using TripleSpec \citep{herter08} at the 5.2 m Hale Telescope at Palomar Observatory. The spectrograph covers from 1.0--2.4 µm with a spectral resolution of $\sim$2500 with a $1^{\prime\prime}$ slit. The spectra were obtained at airmass $\sim 1\farcs1$ with four 80 s exposures in an ABBA sequence.
%
The nearby A0V star HD197291 was observed at similar airmass for telluric correction and flux calibration.  The spectra taken with TripleSpec were processed using SPEXTOOL package version 4.1 \citep{vacca03,cushing04}, which includes pre-processing, aperture extraction, and wavelength calibrations.  The individual wavelegth calibrated spectra are median combined using  XCOMBSPEC in SPEXTOOL. Telluric correction and flux calbration are applied using XTELLCOR to produce the final calibrated spectrum. 

HBC 722 was observed on WHT with the low-resolution optical spectrograph ISIS on 2012 August 9 in an observing run focused on disks with inner cavities, as identified from mid-IR spectral energy distributions.  The data reduction, including flux calibration from spectrophotometric standards, is described in  \cite{vandermarel16}.  The spectrum was obtained in August 2012, during a faint period following the initial burst \citep{kospal16}.  We multiply the flux by 2.2 to match the TripleSpec spectrum.  This scale factor is roughly consistent with the increased brightness between 2012--2016.

In addition to spectroscopy, we also obtained archival photometry (Table \ref{tab photometry}) for the three objects.  FU Ori has gotten fainter by 0.5 mag at optical wavelengths between the 2MASS observation and 2022, so some offset is expected (see photometry in \citealt{herbst99} and from the ASAS-SN survey of \citealt{shappee14} and \citealt{kochanek17}).  
{V883 has shown a $\sim1$~mag range in {\it Gaia} $G$ band \citep{gaiaDR3var}.}
HBC 722 underwent a dip in brightness in late 2011, several months after its outburst, and it remains bright henceforth \citep{kospal16}. Since our goal with these data were to help validate and guide our models, we did not attempt to scale these datasets to any single epoch.


\begin{figure*}
    \centering
    \vspace{-1.5em}
    \includegraphics[width=\textwidth]{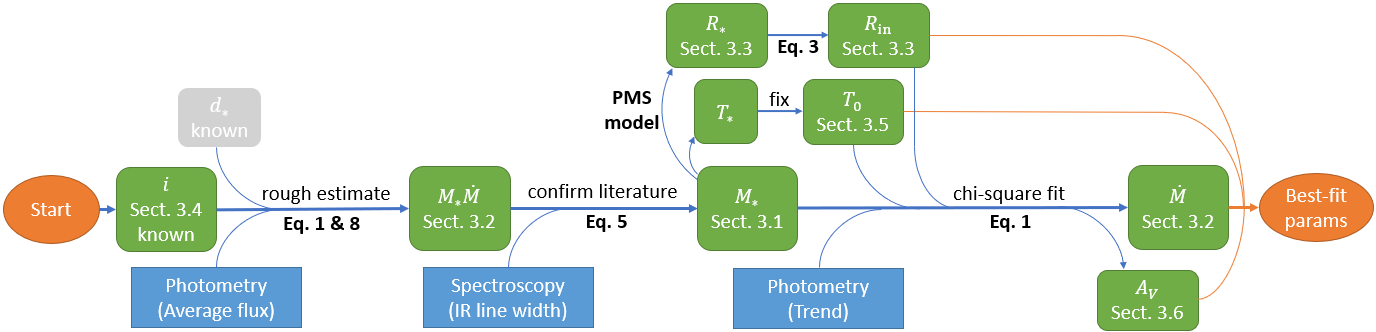}
    \caption{{Flowchart showing the process of FUor fits. Parameters summarized in Table \ref{tab fiducial} are in green rounded rectangles, and spectrophotometric input data in blue boxes. Only the photometry is quantitatively used for the $\chi^2$ fitting. Equations and models relevant to each step are annotated in bold. Blue arrows indicate processes, and orange ones on the right are collection of best-fit parameters. The stellar distance $d_*$ is only adopted from previous studies but not examined here, hence shown in grey.}}
    \label{fig flowchart}
\end{figure*}

\begin{figure*}[ht]
\vspace{-3.0em}
    \centering
    \includegraphics[scale=0.52]{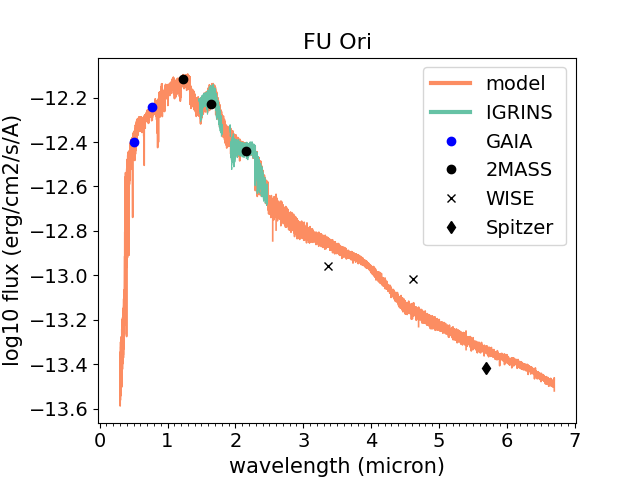}
    \includegraphics[scale=0.52]{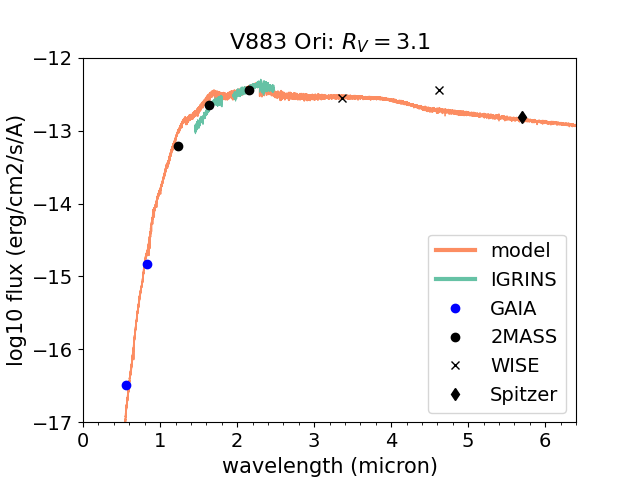}
    \includegraphics[scale=0.52]{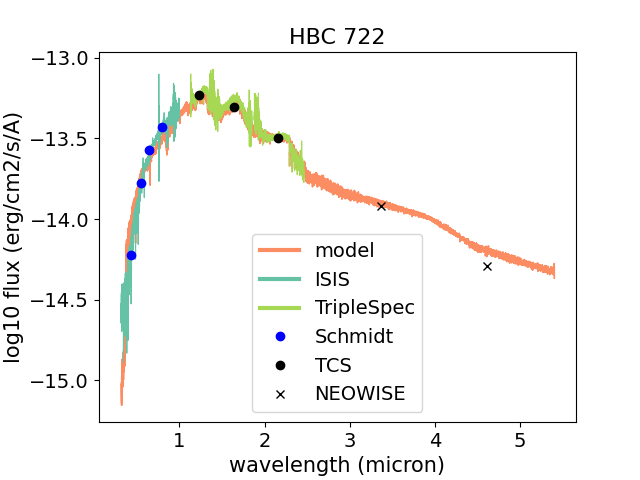}
        \includegraphics[scale=0.52]{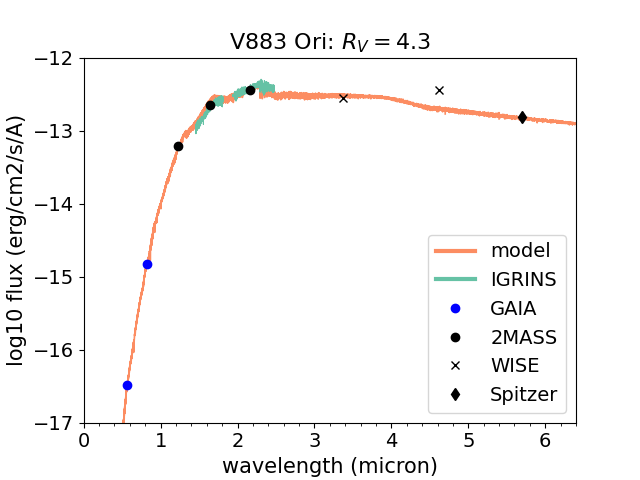}
    \caption{Observed data and best-fit model of the SEDs of three FUors: FU Ori (upper left), HBC 722 (bottom left) and V883 Ori ($R_V=3.1$, upper right and $R_V=4.3$, lower right; see text in Sect.~\ref{fits to data}).}
    \label{Fig FU_Ori overview}
    \vspace{1em}
\end{figure*}

\begin{deluxetable*}{ccccccccc}[ht]
\label{tab fit}
\tablecaption{Parameters of best-fit models for FUor objects}
\tablehead{\colhead{Target} & \colhead{$i$} & \colhead{$d_*$} & \colhead{$M_*$} &\colhead{$\dot{M}$}  & \colhead{$R_*$\tablenotemark{$\dagger$}}& \colhead{$T_0$} & \colhead{$A_V$} & \colhead{$L_{\rm bol, total}$} \\ \colhead{} & \colhead{} & \colhead{(pc)} & \colhead{(${\rm M_\odot}$)} & \colhead{(${\rm M_\odot}$/yr)} & \colhead{(${\rm R_\odot}$)} & \colhead{(K)} & \colhead{(mag)} & \colhead{(${\rm L_\odot}$)} } 
\startdata
FU Ori & $37^{\circ}$ (1) & 408 (2) & 0.6 & $2.8\times10^{-5}$ & 2.0 & 4000 & 2.1 & 119\\
V883 Ori\tablenotemark{$\ddagger$} & $38^{\circ}$ (3) & 388 (4) & 1.3 & $1.1\times10^{-4}$ & 2.7 & 4600 & 15.0 & 647\\
HBC 722 & $73^{\circ}$ (5) & 795 (6) & 0.6 & $2.7\times10^{-5}$ & 2.0 & 4000 & 3.1 & 115
\enddata
\tablenotetext{\dag}{In all three cases, $\dot{M}$ is sufficiently high that $R_{\rm in}=R_*$.}
\tablenotetext{\ddag}{Best fit with $R_V=4.3$.}
\tablerefs{(1) \cite{labdon20}; 
(2) \cite{bailerjones18}; (3) \cite{cieza16}; (4) \cite{kounkel17}, \citet{lee19}; (5) \cite{kospal16}; (6) \cite{kuhn20}}
\vspace{-1.5em}
\end{deluxetable*}

\subsection{Source properties from fits to data}
\label{fits to data}


{We show in Fig.~\ref{fig flowchart} a schematic of our fitting process for the three FUors described above. Since we only intend to validate our models in this section, we provide 
a simple workflow applicable where the viscous disk dominates. 
First, we use Eq.~\ref{eq Tviscous} and \ref{eq viscous flux} to roughly estimate $M_*\dot{M}$ from the average flux of the photometry of a given target combined with the inclination (Sect.~\ref{inclination angle}) and the distance inferred by previous studies. 
This allows us to synthesize model spectra with spectral lines reasonably close to the observation. Then, we select metal and molecular absorption lines in the $H$- and $K$-band that clearly reflect rotational broadening (e.g., \ion{Fe}{1} at 1530 nm, \ion{Al}{1} at 1675 nm and the CO overtone bandhead at 2293 nm; more in Fig.~\ref{Fig FU_Ori} and \ref{Fig V883_Ori}). We compare model and observed line widths by visual inspection to confirm literature estimates of the central mass $M_*$ (cf. Eq.~\ref{eq broadening}).

With $M_*$ available, we obtain $R_*$ and $T_*$ from the PMS model (Sect.~\ref{stellar photosphere}). \ri\ is then ready from Eq.~\ref{eq rin}. The free parameter $T_0$ at best has a mild effect in improving the fit in the fast accretion case (Sect.~\ref{tzero}), and we thus fix $T_0=T_*$, with the warning that slower accretion rates or longer wavelengths may render variation of $T_0$ important. 

We then perform a $\chi^2$ fit on our synthetic SED to determine $\dot{M}$ and $A_V$ simultaneously.} To measure the discrepancy between the model and the observation, we calculate the $\chi^2$ values on the observed photometric magnitudes $\rm mag_{obs}$ and the corresponding model magnitudes $\rm mag_{exp}$ in specific photometric bands:
\begin{equation}
    \chi^2 = \rm \frac{(mag_{obs} - mag_{exp})^2}{(0.1mag)^2}
\end{equation}
We apply filter profiles from the SVO Filter Profile Service \citep{svo20} to calculate the magnitudes and assume that the observational error is uniformly 0.1 mag to ensure equal weighting. We do not fit the SED based on the spectra {in Table \ref{tab spectroscopy}}, which are not available through all wavelengths of interest here.

\begin{figure}
    \centering
    \includegraphics[width=0.47\textwidth]{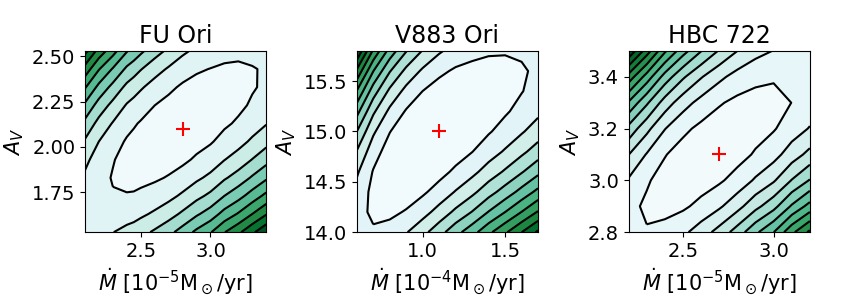}
    \includegraphics[width=0.47\textwidth]{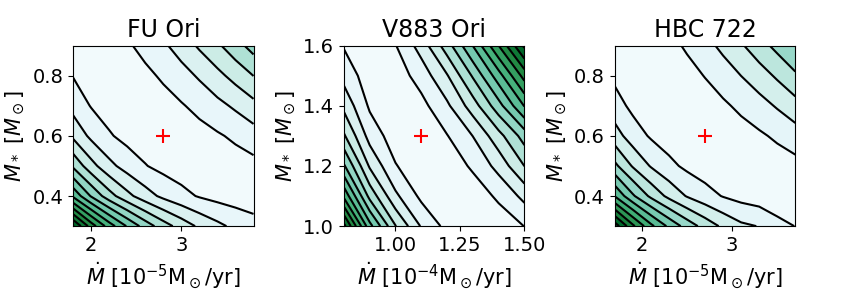}
    \caption{{Contour of $\chi^2$ values of each fit. Dark green means large $\chi^2$. Red plus signs indicate the best fit locations. {\it Top:} Degeneracy between $\dot{M}$ and $A_V$, with other parameters invariant as in Table \ref{tab fit}. {\it Bottom:} Degeneracy between $\dot{M}$ and $M_*$. $R_*$ and $T_0$ (kept equal to $T_*$) are changed along with $M_*$ via the PMS model.}}
    \label{fig degeneracy}
\end{figure}

\begin{figure*}[ht]
    \centering
    \includegraphics[scale=0.35]{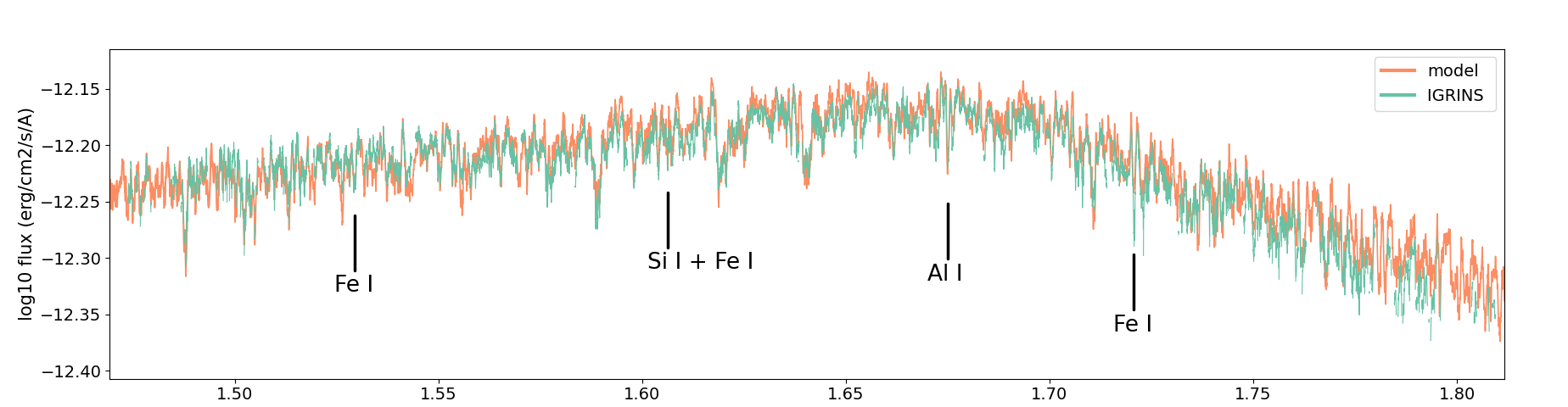}
    \includegraphics[scale=0.35]{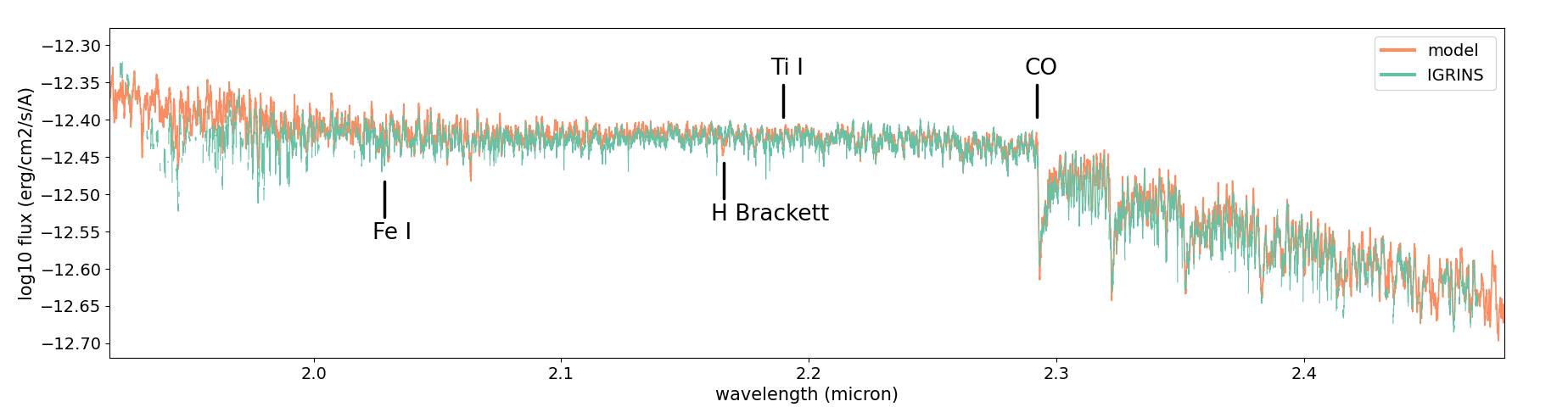}
    \includegraphics[scale=0.25]{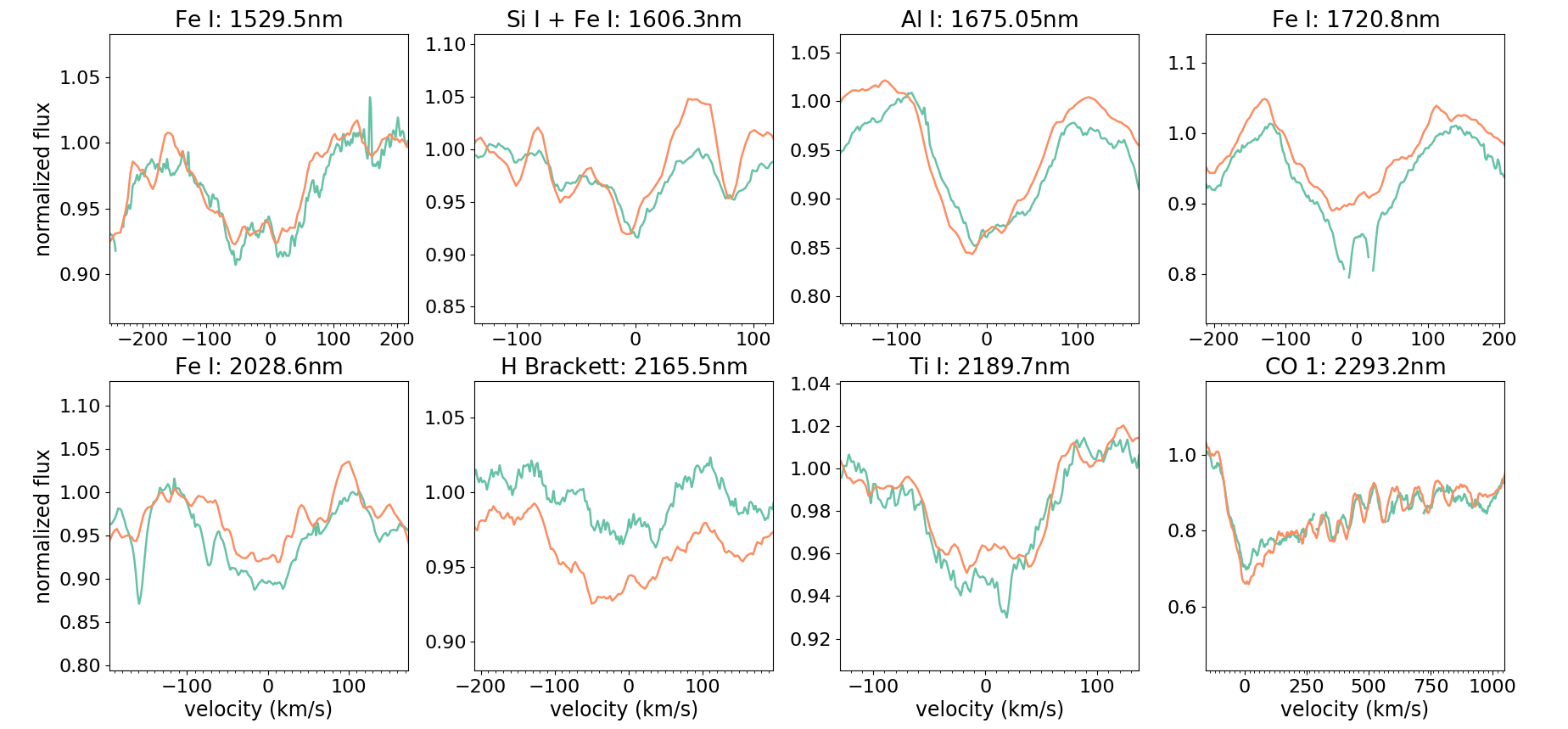}
    \caption{Model and observed spectra of FU Ori in the $H$- and $K$-band, along with 8 selected lines.}
    \label{Fig FU_Ori}
    \vspace{1em}
\end{figure*}

\begin{figure*}[ht]
    \centering
    \includegraphics[scale=0.35]{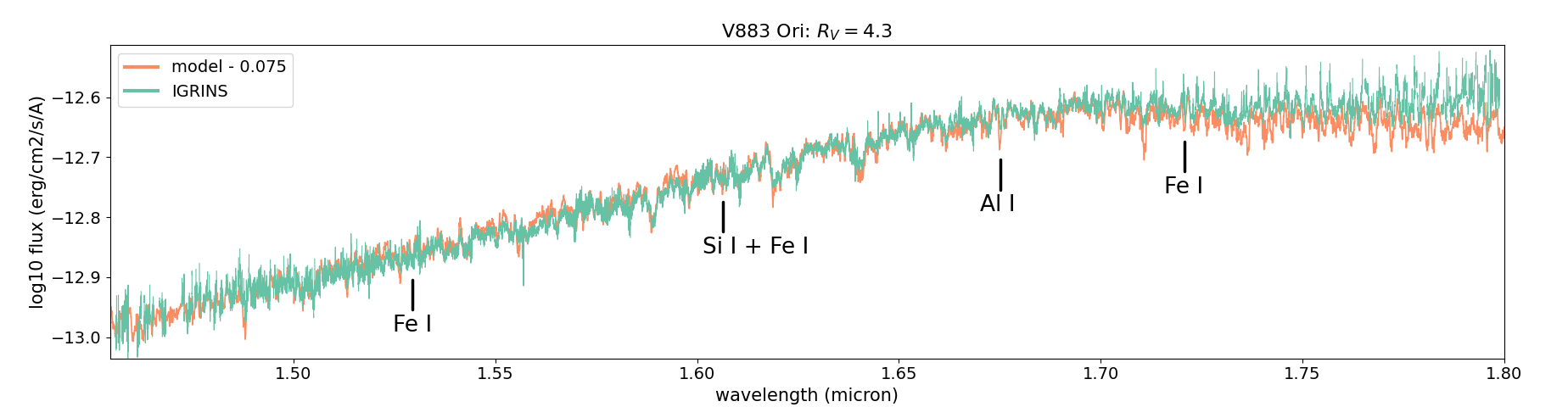}
    \includegraphics[scale=0.256]{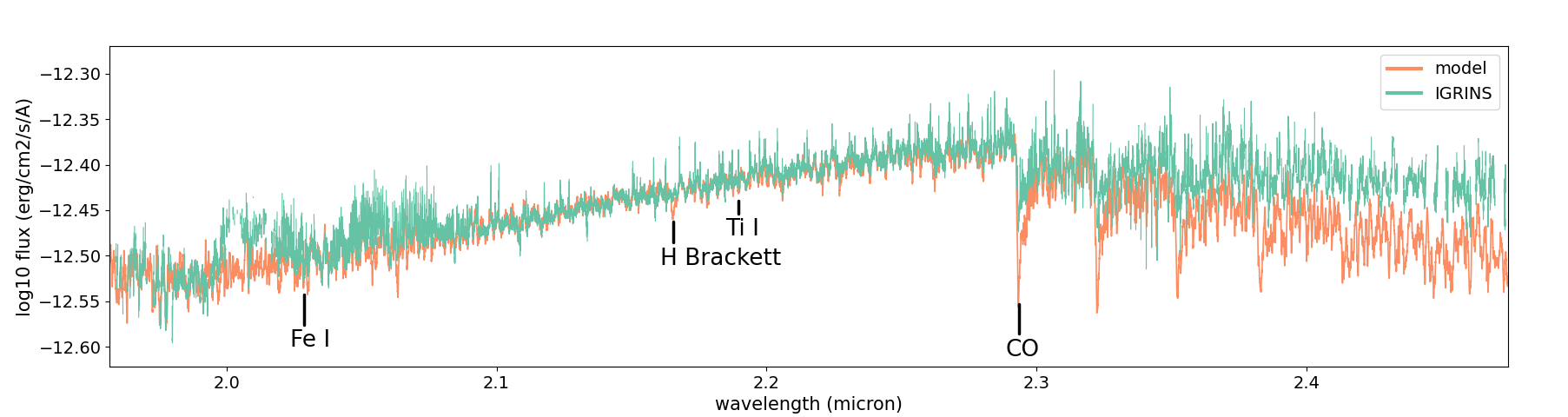}
    \includegraphics[scale=0.35]{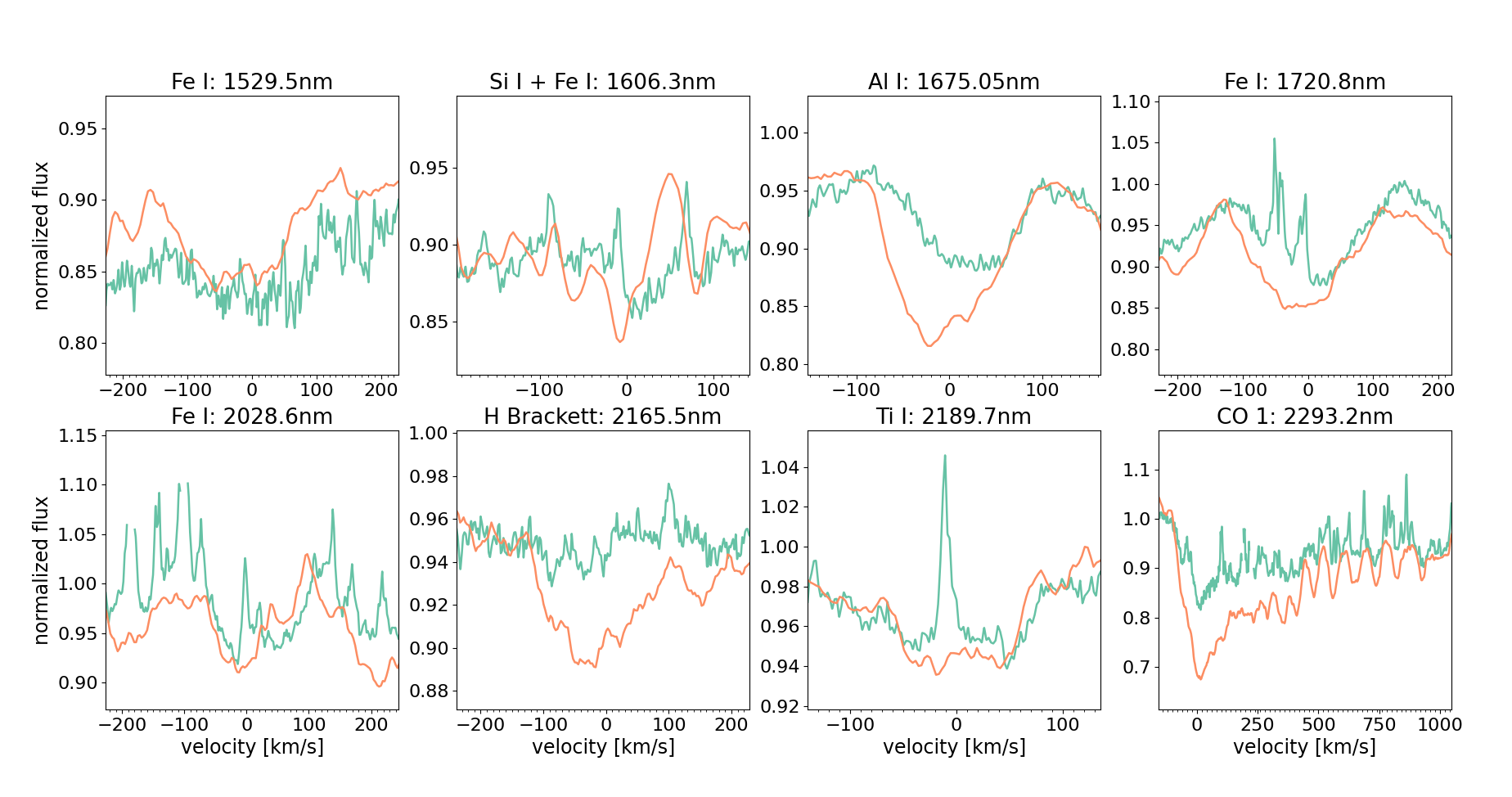}
    \caption{Model and observed spectra of V883 Ori in the $H$- and $K$-band, along with 8 selected lines. {The $H$-band model spectrum is shifted down by 0.075.}}
    \label{Fig V883_Ori}
    \vspace{1em}
\end{figure*}

Table~\ref{tab fit} shows the parameters and results for the fits, as well as the dereddened model bolometric luminosity of all emission components $L_{\rm bol, total}$ for each target. Fig.~\ref{Fig FU_Ori overview} displays our best-fit models compared with the data, {with $\chi^2$ contours plotted in Fig.~\ref{fig degeneracy}.} We also demonstrate $H$- and $K$-band details of model and IGRINS spectroscopy for FU Ori and V883 Ori in Fig.~\ref{Fig FU_Ori} and Fig.~\ref{Fig V883_Ori} respectively. 

The broadband fits are affected by the extinction law adopted for our analysis.  FU Ori is relatively mildly embedded, making appropriate our assumption of $R_V=3.1$ in the extinction model. V883 Ori, on the other hand, is heavily embedded in an envelope, with larger grain sizes that lead to grayer extinction law at near-IR and visible wavelengths. Therefore, we show in Fig. \ref{Fig FU_Ori overview} two versions of model V883 Ori spectra: one with fixed $R_V=3.1$ for consistency with other fits, the other with $R_V=4.3$. The latter version can better reproduce the trend of the optical photometries and $H$-band spectrum, and we thus adopt the fit parameters of V883 Ori with $R_V=4.3$ in Table \ref{tab fit} and Fig.~\ref{Fig V883_Ori}.

{The quantified fit to obtain $\dot{M}$ is subject to uncertainties both from model degeneracy and observational limitations. Our workflow circumvents the degeneracy from an uncertain inclination (Sect.~\ref{inclination angle}). However, at short wavelengths without observed spectroscopy, the limited photometric coverage leaves out high-order SED trends that help to solve the degeneracy between accretion rate and extinction. Indeed, contour plots in Fig.~\ref{fig degeneracy} ({\it top}) show that the $\chi^2$ minimum for each fit extends diagonally, as discussed in Sect.~\ref{extinction}. Furthermore, detailed comparisons to past measurements and historical photometry reveal real changes of accretion rate versus time.

We have also attempted to fit $M_*$, but the parameter is very poorly constrained by photometry itself, although the $\chi^2$ values imply a precise estimate of $M_*\dot{M}$ (Fig.~\ref{fig degeneracy}, {\it bottom}). Discrepancies between template and real spectral profiles also dominate over line width differences caused by the stellar mass, making it difficult in practice to fit $M_*$ with spectroscopy. Detailed modeling of individual lines may provide more reasonable constraints of $M_*$ but would then go beyond our purpose of model validation.}



Our measured accretion rates are consistent with accretion rates measured previously for these three objects (Table \ref{tab accretion reference}).  We discuss the fits to the three sources in the following sub-subsections. {As remarked above, time-domain variability introduces significant uncertainties to accretion rates.    
The methodological differences and geometrical assumptions in converting empirical measurements to accretion rates also dominate over formal uncertainties.}

\begin{deluxetable}{ccccc}[ht]
\label{tab accretion reference}
\tablecaption{Comparison of best-fit $\dot{M}$ for FUor objects}
\tablehead{\colhead{Target} & \colhead{$M_*$} & \colhead{$\dot{M}$} & \colhead{$T_{\rm max}$} & \colhead{Reference} \\ \colhead{} & \colhead{(${\rm M_\odot}$)} & \colhead{(${\rm M_\odot}$/yr)} & \colhead{(K)} & \colhead{} } 
\startdata
FU Ori & 0.6 & $3.8\times10^{-5}$ & 6300\tablenotemark{\footnotesize $a$} & (1)\\
       & 0.6 & $2.8\times10^{-5}$ & 8800 & this work\\
V883 Ori\tablenotemark{$\dagger$} & 1.3 & $7\times10^{-5}$ & 10800\tablenotemark{\footnotesize $b$} & (2)\\
         & 1.3 & ${1.1}\times10^{-4}$ & 12100 & this work\\
HBC 722 & 0.66 & $1.3\times10^{-5}$ & 7100\tablenotemark{\footnotesize $c$} & (3)
\\
        & 0.6 & $2.7\times10^{-5}$ & 8800 & this work
\enddata
\tablenotemark{$\dagger$}{Best fit with $R_V=4.3$.}
\tablenotetext{\footnotesize a}{We cite $R_{\rm in}=3.5\ {\rm R_\odot}$ from the reference and calculate $T_{\rm max}$ from Eq.~\ref{eq Tviscous}.}
\tablenotetext{\footnotesize b}{As no $R_{\rm in}$ is provided in the reference, we calculate $T_{\rm max}$ with $R_*$ from the PMS model.}
\tablenotetext{\footnotesize c}{The reference provides the best-fit $T_{\rm max}$ defined in the same way as this work.}
\tablerefs{(1) \cite{perez20} updated from $2.4\times10^{-4}$~M$_\odot$ yr$^{-1}$ of \cite{zhu07}; (2) \cite{cieza16}; (3) \cite{rodriguez22}}
\vspace{-3.0em}
\end{deluxetable}

\subsubsection{FU Ori} {Our fit to FU Ori agrees with both the photometry and spectroscopy, although only the SED has been used for fitting.} The best-fit accretion rate is obtained $2014$, with a $B$-band magnitude of 11.0 \citep{siwak18}; our best-fit has $B=10.8$.  The peak of the light curve, in the year $\sim 1937$, had $B=9.6$ \citep{hoffleit39}, which corresponds to an accretion rate of $1.0\times10^{-4}~{\rm M_\odot}$/yr, or $3.6$ times higher than the present-day accretion rate.  


The accretion rate calculated here for FU Ori is similar to that from \citet{perez20}, despite our {higher} $T_{\rm max}$. \citet{perez20} adopted the $T_{\rm max}$ measured by \citet{zhu07} but obtained an accretion rate an order of magnitude lower.  The lower disk inclination in \citet{perez20} led to a larger $M_*$ for the same line width (Eq.~\ref{eq broadening}), a smaller $M_*\dot{M}$ due to a dimmer true luminosity, and a smaller $R_{\rm in}$ (Eq.~3 in \citealt{zhu07}, where $R_{\rm in}$ is constrained in a different way from here).  The higher $T_{\rm max}$ measured here is consistent with the ultraviolet measurements from \citet{kravtsova07}.

However, the updated $R_{\rm in}=3.5~{\rm R_\odot}$ from \citet{perez20} is still larger than the value $R_{\rm in}=2.0~{\rm R_\odot}$ here, which accounts for the difference in $T_{\rm max}$. It has been suggested that the large inner disk radius is probably a result of expansion of the central star at high $\dot{M}$ \citep[see discussion in][]{hartmann85}. This picture is not included in our model setting, which we identify as an important source of uncertainty in our methods. 

\subsubsection{V883 Ori} 
\label{v883ori}
V883 Ori is accreting at a rate of  $1.1\times10^{-4}$ M$_\odot$ yr$^{-1}$, the strongest of the three stars presented here.  Although technically FUor-like because the burst was not detected, V883 Ori is widely considered a bona fide FUor object based on the spectral characteristics (first classified as FUor-like by \citealt{strom93}; see also \citealt{reipurth97,connelley18}).  
{The $\sim1$~mag range in {\it Gaia} $G$ band of V883 Ori corresponds to a factor of 1.6 in accretion rate uncertainties.}

\citet{cieza16} arrived at $M_*=(1.3\pm0.1)~\rm M_\odot$ for V883 Ori via dynamic mass measurement and then estimated the accretion rate with $M_*$ and the bolometric luminosity. While no formal uncertainty is available, the bolometric luminosity value is subject to errors of the inclination and the extinction, leading to significant uncertainties in accretion rate, as described above for FU Ori.

While Fig.~\ref{Fig FU_Ori} demonstrates good agreement between model and observation for FU Ori in the $H$- and $K$-band, V883 Ori's near-IR spectrum (Fig.~\ref{Fig V883_Ori}) probably consists of emission features that appear as spikes in \ion{Fe}{1} and \ion{Ti}{1} lines and shallower \ion{Al}{1}, H Brackett and CO lines. Gravity and metallicity that deviate from our assumptions may also play a role. {The WISE photometry of V883 Ori is heavily saturated, leading to large uncertainties.  The NEOWISE photometry \citep{mainzer14} spans over 1.3 mag, likely because measurements of such a saturated object are challenging.  The $W_2$ band may also include contributions from CO and H$_2$ emission, perhaps in outflow cavity walls or bow shocks \citep[see, e.g., extended green objects][]{marston04,cyganowski09}, although usually the outflow emission is much fainter than what is observed to V883 Ori.}


\subsubsection{HBC 722} 

\begin{figure}
    \centering
    \includegraphics[scale=0.5]{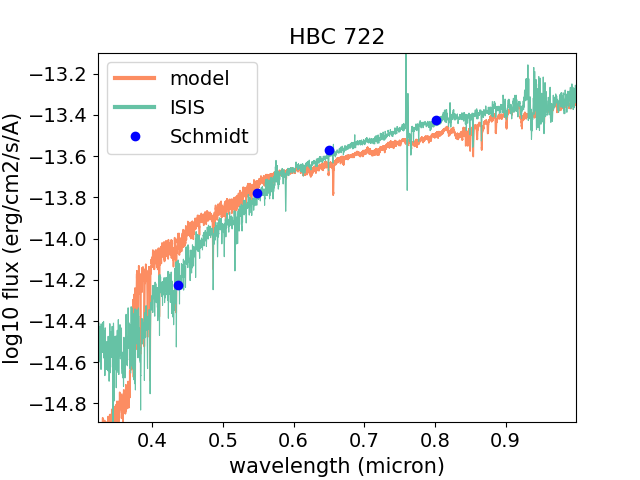}
    \caption{{Observed data and best-fit model of the SEDs of HBC 722 in the optical. This is an enlargement of Fig.~\ref{Fig FU_Ori overview} (bottom left).}}
    \label{fig HBC722_fit(optical)}
\end{figure}

For HBC 722, the accretion rate is lower than for V883 Ori and FU Ori.  We use the photometry taken when the lightcurve of the object reached a steady bright state in May 2015, yielding an accretion rate of $2.7\times10^{-5}~ {\rm M_\odot}/$yr. In late 2011 and early 2012, when the $I$ band was dimmer by 1.3 mag, a lower value of $\dot{M} = 8\times10^{-6} {\rm M_\odot}/$yr is needed for the fit. Our estimate of $\dot{M}$ is higher than the best fit from \citet{rodriguez22} ($\log\dot{M}=-4.90_{-0.40}^{+0.99}~\rm M_\odot/yr$), which can be explained by the variability of the source and the different inclination angles.  A lower $\dot{M}$ in our model will also bring our $T_{\rm max}$ closer to the cited work ($T_{\rm max}=7100_{-500}^{+300}$~K). The model best fit of $R_*=2.07\ {\rm R_\odot}$ from \citet{rodriguez22} also suggests that bloating of the central star is probably not significant for HBC 722.

{The high $T_{\rm max}$ is likely responsible for the bluer optical SED of our best-fit spectrum than observation, shown in Fig.~\ref{fig HBC722_fit(optical)}. An alternative explanation is that the extinction law has a smaller $R_V$, the opposite to what is expected from Galactic star-forming regions. Our model spectrum also presents strong absorption at the Balmer jump at $365$~nm, a continuum feature commonly observed among late-B to early-F main-sequence stars and in excess emission in accreting T Tauri stars \citep{gullbring98}. We cannot remove the feature, which comes from the viscous disk where $T_{\rm visc} \gtrsim 6000$~K, without severely downsizing the SED.   Neither HBC 722, shown here, nor V1057 Cyg presented in \citet{valenti93} have a strong Balmer Jump.}



\section{Observational Diagnostics of FUors}
\label{observational diagnostics}

\begin{figure*}[ht]
    \centering
    \includegraphics[scale=0.9]{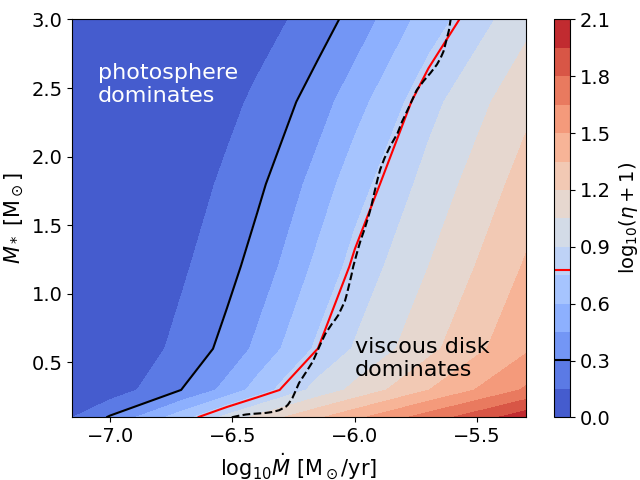}
    \caption{Comparison of {summed $H$- and $K$-band fluxes} between the viscous disk and the photosphere with different stellar masses and accretion rates. Filled colors indicate the ratio of flux of the viscous disk to that of the photosphere ($\eta$) for a given mass in a logarithmic scale. The black solid line (``transition line") shows where the two fluxes become equal ($\eta = 1$); the red line (``sufficient dominance line") shows where $\eta = 5$. The black dashed line shows where $R_{\rm in}=R_*$. Other parameters are held constant at $i = 45^\circ, A_V=10$ mag, and $T_0$ as the minimal value $T_*$ given by the PMS model.}
    \label{fig acc_compare}
    \vspace{1em}
\end{figure*}

FUor events are classically defined as strong (factors of tens to hundreds) increases in luminosity that last for decades to centuries.   These outbursts display unique spectral characteristics in the infrared, which informs identification and classification of FUors among YSO candidates. \citet{connelley18} (hereafter CR18) summarized salient spectral criteria in the near-IR, including strong CO absorption, water vapor bands (``triangular" $H$-band continuum), and weak metallic lines.  In this section, we analyze outburst behaviors and characteristic FUor features in the context of our star-disk model and discuss implications on low-luminosity FUors.

\subsection{Viscous disk versus photosphere}
\label{viscous disk versus photosphere}

According to our discussion in Sect.~\ref{accretion rate}, the accretion rate has a significant bearing on the luminosity of the viscous disk.  If the accretion rate is low, then the stellar photosphere is the main source of luminosity in the near-IR, while the passively heated disk dominates at longer wavelengths.  

To investigate the link between physical properties and observational constraints of low-luminosity FUors, we describe how the contributions from the stellar photosphere and viscous disk vary with $\dot{M}$. {We introduce a parameter
\begin{equation}
    \eta \equiv \frac{F_{\rm visc, H} + F_{\rm visc, K}}{F_{\rm phot, H} + F_{\rm phot, K}},
\end{equation} the ratio of the summed $H$- and $K$-band flux} of the viscous disk to that of the photosphere. As Fig.~\ref{fig acc_compare} demonstrates, the turning point (``transition line", black solid, where $\eta = 1$) occurs at $2 \times 10^{-7}~{\rm M_\odot}/$yr for $M_* = 0.3~{\rm M_\odot}$; the $\dot{M}$ value of this line varies with the stellar mass. On the left of the transition line, if $\dot{M}$ is so low that the maximal temperature of the inner disk falls below 1400~K, the source will have no viscous disk component, hence $\eta=0$ (dark blue region in Fig.~\ref{fig acc_compare}) and our model degenerates into a TTS. This occurs where $\dot{M}\lesssim10^{-7}~{\rm M_\odot}/$yr. On the right of the transition line, we also plot the ``sufficient dominance line" (red solid), where we take $\eta = 5$, to indicate where the viscous disk sufficiently outshines the photosphere so that the spectrum features a typical viscous disk.

Before further explaining the value choice of $\eta$ in defining the sufficient dominance line in the following subsections, we note that this line concurs with the black dashed line that shows where $R_{\rm in}=R_*$ and the magnetospheric accretion ceases to exist. {However, if the bolometric flux is used to define $\eta$, then the $\eta$ contour will shift leftwards by $0.07$ as the viscous disk, if any, is almost always brighter than the stellar photosphere in the mid- to far-IR.}

\subsection{Outbursts in optical color-magnitude diagrams}
\label{outbursts in optical}

\begin{figure*}
    \centering
    \includegraphics[scale=0.5]{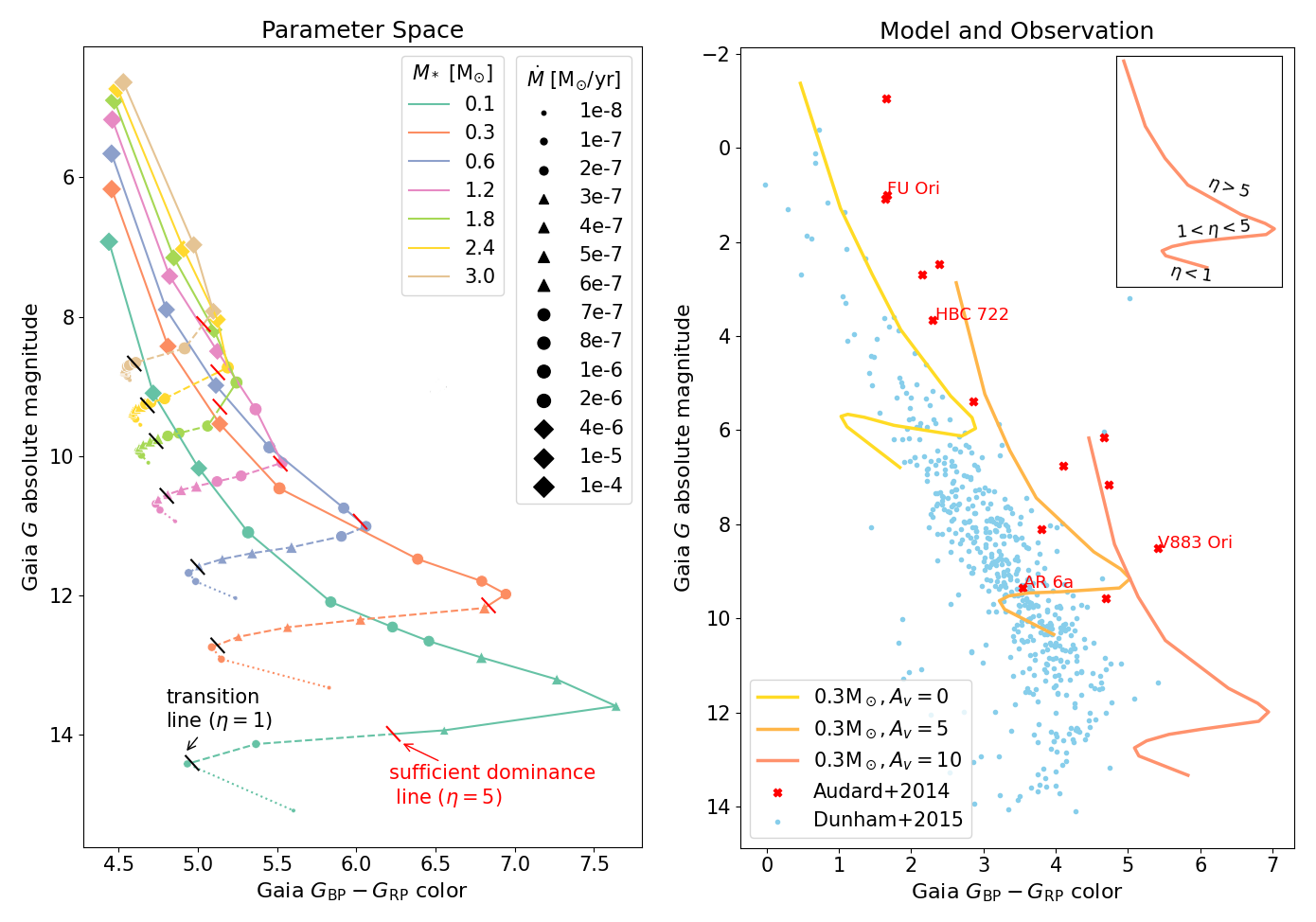}
    \caption{Color-magnitude diagram in the optical. {\it Left: } Model iso-mass curves through the parameter space. Series of points with the same color {but different sizes and markers} represent different masses (from 0.1~${\rm M_\odot}$ to 3.0~${\rm M_\odot}$). Black and red dashes respectively mark the transition line and the sufficient dominance line, which separate the iso-mass curves into three sections: $\eta<1$ (dotted), $1<\eta<5$ (dashed) and $\eta>5$ (solid). These three sections agree with the apparent segments of the curves except for the sufficient dominance at $M_*=0.1~{\rm M_\odot}$. Extinction is set at $A_V=10$ mag. {\it Right: } Comparison of model and observation. $M_*=0.3~{\rm M_\odot}$ is fixed for the iso-mass curves, while extinction varies from 0 to 10. Photometry from \citet{audard14} (FUors; red crosses) and \citet{dunham15} (YSOs; blue dots) are scaled to $d_* = 10$ pc (see text). 
    Names of FUors in Sect.~\ref{fits to data} {plus those specifically discussed in Sect.~\ref{outbursts in optical}} are annotated. The in-box plot is an overview of the $M_*=0.3~{\rm M_\odot}, A_V=10$ curve, with three segments corresponding to three $\eta$ intervals.}
    \label{fig color-magnitude_gaia}
    \vspace{1em}
\end{figure*}

Our model opens the prospect of low-luminosity FUor candidate identification in all-sky surveys at optical or near-IR wavelengths. Plotting color-magnitude diagrams (CMDs) in optical and infrared colors, we may delineate potentially fruitful regions to search for unidentified FUors. 

In Fig.~\ref{fig color-magnitude_gaia}, we display the CMD at optical wavelengths with {\it Gaia} EDR3 \citep{GAIA} colors and magnitudes. Through the parameter space ({\it left}), all iso-mass curves in the figure follow zigzag trails: for a given stellar mass, as the accretion rate grows from $1\times10^{-8}$ to $1\times10^{-4}~{\rm M_\odot}/$yr, the curve first mildly goes up to the left, then turns rightwards, before turning once again to the up-left direction. This shape is mainly due to the  magnetospheric accretion (Eq.~\ref{eq mag}), approximated as a $8000$~K blackbody component. An increase of $\dot{M}$ accompanies a shrinkage of $R_{\rm in}$, causing $L_{\rm mag}$ to peak before dwindling towards zero. The retrograde segment results from the rapid decline of $L_{\rm mag}$ as $R_{\rm in}$ approaches $R_*$. This trend ends where the magnetospheric accretion vanishes; therefore, the second turning points match the sufficient dominance line, where a YSO fully develops into an FUor-like object with a predominant viscous disk. For $M_* = 0.1~\rm{M_\odot}$, the sufficient dominance line is lower than the second turn, which can be explained by
the higher required accretion rate than $\eta=5$ to push the truncation radius to the stellar surface (bottom of Fig.~\ref{fig acc_compare}, where the red line and the black dashed line are separated).

After the second turn, with the heating-up and expansion of the viscous disk, the now-FUor-like object follows a familiar CMD trend, brightening as the color becomes bluer. While in the previous stage, the increment of $\dot{M}$ does not lead to a prominent rise in G-band luminosity, a further growth in accretion rate can readily be detected as an optical outburst.  While the details of these curves have some uncertainty, the general trends are robust, depending only on the fact that $L_{\rm mag}$ first increases and then decreases with $\dot{M}$. 


We compare these model results with available FUor and YSO data sets in Fig.~\ref{fig color-magnitude_gaia} ({\it right}) for FUor and FUor-like objects in the collection of eruptive young stars by  \citet{audard14}.  We also plot YSOs in the {\it Spitzer} catalog of YSOs in the Gould Belt from \citet{dunham15} with {\it Gaia} EDR3 colors, excluding targets in 
targets in Aquila, Lupus {\uppercase\expandafter{\romannumeral5}}, Lupus {\uppercase\expandafter{\romannumeral6}} and Serpens to avoid contamination by background Asymptotic Giant Branch (AGB) stars. 
We update the molecular cloud distances with {\it Gaia} DR2 parallax where available \citep{dzib18}. A total number of 14 FUor-type objects and 526 YSOs are presented .

In Fig.~\ref{fig color-magnitude_gaia} ({\it right}),  an iso-mass curve of $M_*=0.3~{\rm M_\odot}$ with varied extinction that spans from $A_V=0$ to $A_V=10$ are consistent with the data. {The only member in the \cite{audard14} collection far below the sufficient dominance line is AR 6a, which had a $K$-band spectrum in 2002--2003 with strong CO and Br$\gamma$ absorption, characteristic of FUors \citep{aspin03}. However, a more recent spectrum taken by \cite{connelley18} indicate a change in the spectrum, which now shows weak CO absorption and resembles a reddened early-type star.} Only a small extinction range suffices to cover the data, as higher extinction challenges the sensitivity of {\it Gaia} detection. The agreement suggests feasibility to select mildly extincted targets of interest in search of FUor-like objects: a YSO close to the $\eta>5$ branch (thus similar to an FUor in optical colors and magnitudes) possibly has a dominant viscous disk.  A search that extends to the $1<\eta<5$ branch for those with mixed emission would end up including most YSOs shown here.



\subsection{Outbursts in the mid-IR}
\label{outbursts in the mid-IR}
\begin{figure*}
    \centering
    \includegraphics[scale=0.5]{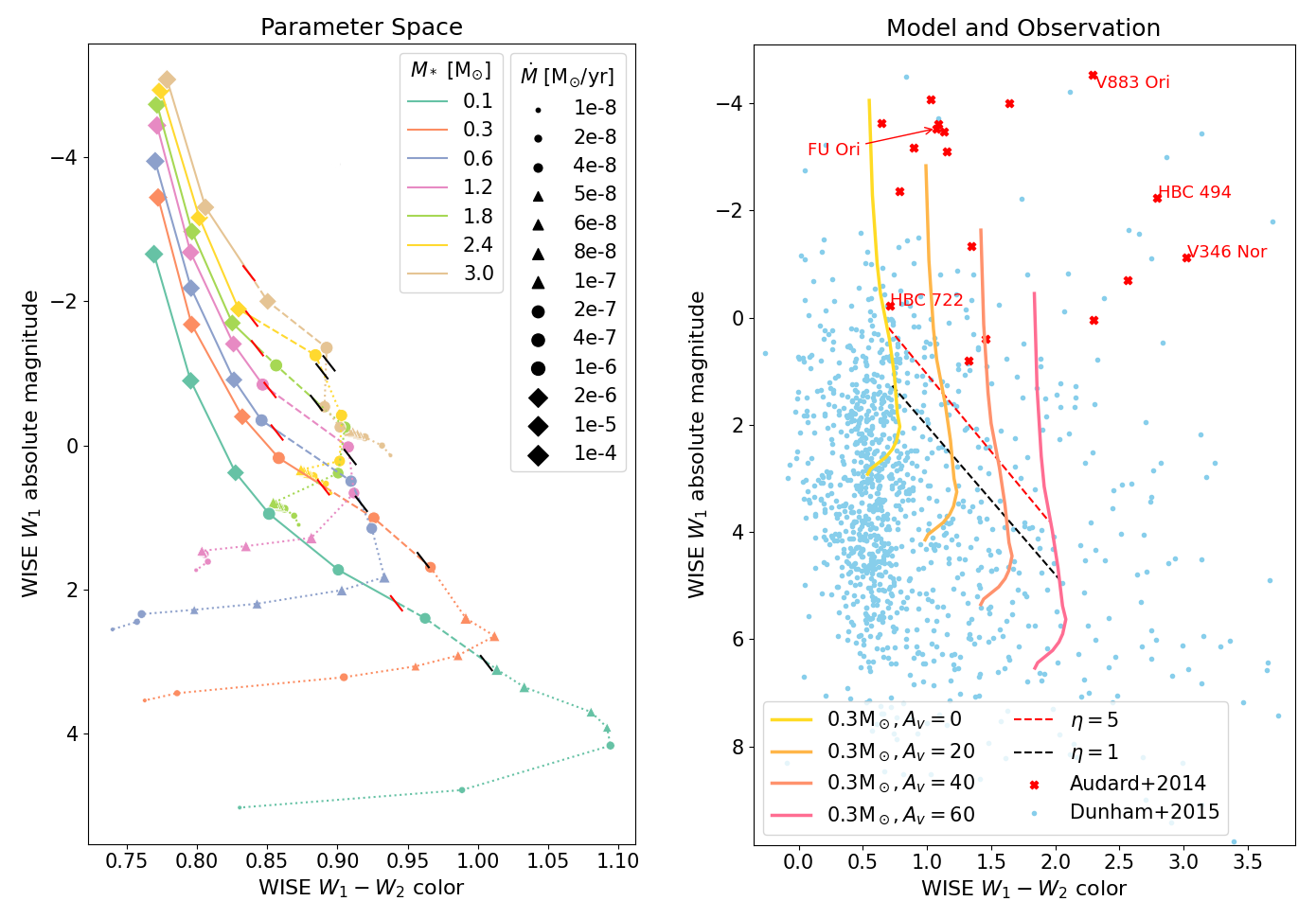}
    \caption{Mid-IR CMD with WISE $W_1$ and $W_2$ colors. {In the {\it right} panel, }black and red dashed lines show the transition line ($\eta=1$) and the sufficient dominance line ($\eta=5$). Names of FUors in Sect.~\ref{fits to data} plus those specifically discussed in Sect.~\ref{outbursts in the mid-IR} are annotated. For other details, see caption of Fig.~\ref{fig color-magnitude_gaia}.}
    \label{fig color-magnitude_WISE}
    \vspace{1em}
\end{figure*}

While the first examples of FUors, including FU Ori itself, were identified by very large brightness increases in the optical, many YSOs are located in dusty, high-extinction environments and can only be detected at longer wavelengths.
Near- and mid-IR surveys such as VVV and NEOWISE enable detection of outburst events in the IR, thus leading to multiple FUor candidates \citep[e.g.,][]{contreras17,guo20,park21}.
In this subsection, we show how WISE $W_1$ and $W_2$ band magnitudes of a YSO changes during an outburst to {help identify ongoing FUor events.}

The mid-IR patterns (Fig.~\ref{fig color-magnitude_WISE}, {\it left}) show important differences from the pattern in the optical. Overall, while the magnitude changes from $\dot{M}=10^{-8}~{\rm M_\odot}/$yr to $\dot{M}=10^{-4}~{\rm M_\odot}/$yr in the optical and the mid-IR are comparable, the WISE $W_1-W_2$ color varies by
only $\sim0.3$ magnitudes compared to $\sim3$ in the optical. As for the pattern, the 
iso-mass curves in the mid-IR show two main segments with a turning point beneath the transition line. In contrast to the optical, where the magnetospheric accretion impresses the curves, the twofold trend in the mid-IR is mainly a result of disk brightening. At low $\dot{M}$, the stellar photosphere dominates the emission, and the moderate viscous disk temperature contributes redder components; the reverse is true at higher accretion rates and higher disk temperature. Another short segment appears at the low ends of iso-mass curves where $M_*>0.6~{\rm M_\odot}$ because of domination of the accretion-invariant passive disk. The blueward trend in this segment stems from the contraction of $R_{\rm in}$ and thus the addition of disk emission near 1400~K.

The range in mid-IR magnitude in our model is similar to the optical in our model, departing from empirical measurements that FUor lightcurves usually show larger brightness change in the optical than in the mid-IR \citep[e.g.,][]{hillenbrand18,szegedi-elek20}. Multi-component models based on simpler assumptions, in contrast, produce a wavelength dependence of magnitude changes \citep{hillenbrand22}. We identify our dipole magnetic field assumption as the main cause of the deviation. High-order magnetic fields attenuate more quickly with radius, so the balance with accretion pressure occurs at 
smaller $R_{\rm in}$; the additional $\sim10^3$ K emission would make the quiescent disk brighter in the mid-IR and reduce the WISE magnitude variation in an outburst. 

Similar to Sect.~\ref{outbursts in optical}, we compare the model and selected 18 FUor-type and 1129 YSO targets from \citet{audard14} and \citet{dunham15}. We perform the same treatment on distances and avoidance of probably AGB-contaminated regions as before. In the FUor-type data set, we omit Z CMa, a binary source composed of an FUor and a Herbig Be star 
with a remarkable absolute magnitude of $W_1=-9.5$~mag, because the infrared brightness of this source is dominated by 
the Herbig Be component \citep{hinkley13}. The YSO data set here is larger than the optical since heavily embedded objects are still detectable in the mid-IR; however, we omit 5 YSOs with $W_1-W_2<-0.5$~mag, 3 sources with $W_1>10$~mag and 6 sources with $W_1-W_2>4.0$~mag where the parameters fall beyond the predictive scope of our models. 

These FUors and YSOs span a large region in the mid-IR CMD (Fig.~\ref{fig color-magnitude_WISE}). While a considerable range of extinction enables the model curve to cover the data adequately, some FUors are brighter and redder than predicted here \citep[e.g., the three upper-right FUors V883 Ori, HBC 494 and V346 Nor:][]{ruiz-rodriguez17_3,ruiz-rodriguez17_2,kospal21}. {Individual considerations are needed to account for the redness: V883 Ori exhibits singular brightness of $W_2$ band presumably due to saturation bias or line emission (Sect.~\ref{v883ori}), while a smooth rise in WISE fluxes of V346 Nor from $W_1$ to $W_3$ band at 11.6~$\mu$m implies warm continuum emission or very heavy extinction. The four WISE bands of HBC 494 present a continuous rise to longer wavelength and a $W_2$ excess. Regardless, the remarkable color of these objects is probably associated with the circumstellar envelope, which may excite molecular emission by interaction with outflows, add mid-IR continuum emission due to scattered light or warm dust, and/or extinct the emission from the star-disk system. Silicon absorption features at 10~$\mu$m reveal the envelope enshrouding these objects \citep{schutz05}.} 


{Another limitation of our models is the simplistic assumption on the magnetic field strength and geometry; nevertheless,} the upper branches of our model curves, especially where $R_{\rm in}$ approaches $R_*$, are less sensitive to this uncertainty. For the purpose of searching for FUor-like objects, our model provides some guidance for target selection. Specifically, in the mid-IR CMD, we suggest that investigating targets above the transition line or the sufficient dominance line may lead to identification of new candidate {FUor-like objects.}

\begin{figure*}[t]
    \centering
    \includegraphics[scale=0.33]{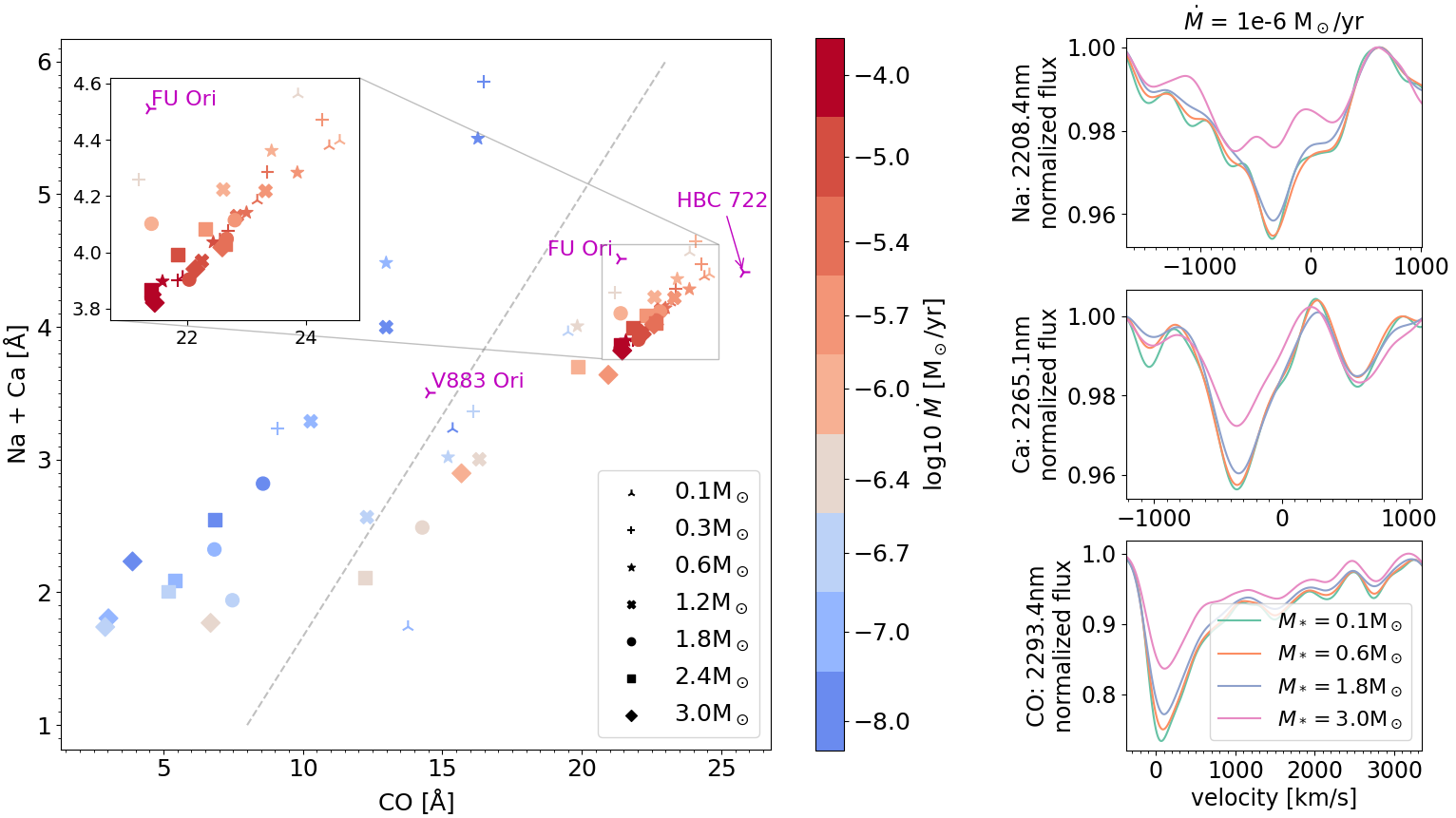}
    \caption{{\it Left: }The EW of CO versus that of Na+Ca for the model objects of various masses and accretion rates and extinction set at $A_V=10$ mag.  Colors represent accretion rates.
    Shapes of points mark different masses. The grey dashed line is the separation of FUors from other YSOs proposed by \citetalias{connelley18}. {Names of FUors in Sect.~\ref{fits to data} are indicated and annotated in magenta.}  The inset panel is an enlargement of  the high-$\dot{M}$ cluster. {\it Right: } Normalized \ion{Na}{1}, \ion{Ca}{1} and CO absorption lines for $\dot{M}=1\times10^{-6}{\rm M_\odot}$/yr {from the multi-component models}.}
    \label{fig equivalent_widths}
    \vspace{1em}
\end{figure*}

\subsection{Equivalent widths of prominent $K$-band features}

Pronounced CO bandheads and weak metallic absorption lines in the near-IR are among the salient characteristics of FUors. In their Fig.~9, \citetalias{connelley18} displayed the observed equivalent widths (EW) of Na+Ca lines ($2.208~\rm \mu m$ and $2.265~\rm \mu m$) versus the CO $v=2-0$ overtone bandhead (from $2.292~\rm \mu m$ to $2.320~\rm \mu m$) of a variety of objects. Bona fide FUors and FUor-like objects lie in a region distinct, with stronger CO lines compared to other YSOs.

The EW diagram serves as a powerful diagnostic of disk physics because (a) the wavelengths of selected lines are close so that the underlying emission is produced at the same radii and have depths that are similarly veiled, 
 and (b) the EW is insensitive to the inclination angle and distance of the source. Comparisons between models and observations with EW diagrams thus focuses our attention to essentials of the source, including the temperature profile, gravity and metallicity.

Varying $M_*$ and $\dot{M}$ with our model, we provide a counterpart plot comparing CO with Na+Ca EWs in Fig.~\ref{fig equivalent_widths}. We convolve our high-resolution model spectra with a Gaussian kernel to match the $R=1200$ resolving power in \citetalias{connelley18} and then measure the EWs by integration of the local area between the spectrum and a baseline tangent to the local continuum\footnote{The CO continuum was estimated by averaging the flux at the short- and long-wavelength end of the $v=2-0$ band.} near the convolved spectral line. As Fig.~\ref{fig equivalent_widths} shows, sources with substantial viscous disk emission (warm-colored) lie along a linear region on the lower-right half of the main panel. This region is separated further into an upper compact cluster and a lower, relatively less densely populated filament. The upper cluster is located at $EW_{\rm CO}\sim22~\rm \AA, EW_{\rm Na+Ca}\sim4~\rm \AA$, with high $\dot{M}$ that corresponds to luminous FUors. The lower filament consists of those with intermediate accretion rates where the viscous disk emission is comparable to that of the photosphere. At the low $\dot{M}$ end, sources with weak or no viscous disk component (cool-colored) correspond to TTSs and follow a distinct linear trend (weaker CO lines), leaving an empty intermediate space between FUors and TTSs. 

\citetalias{connelley18} proposed an empirical separation rule of $EW_{\rm CO}=3\times EW_{\rm Na+Ca}+5~\rm \AA$ (shown as grey dashed line here), consistent with our figure, {although our measurement of $EW_{\rm CO}$ is somewhat smaller overall. V883 Ori crosses the line, probably due to CO emission (see below).} A few low-$\dot{M}$ targets seem to transgress the separation, but they are mostly low-mass YSOs that already have comparatively significant viscous disk component even in low accretion rate. There is only one real exception with no viscous disk component ($M_*=0.1$~M$_\odot, \dot{M}=1\times10^{-8}$~M$_\odot$/yr, at $EW_{\rm CO}=14~{\rm \AA}, EW_{\rm Na+Ca}=3~\rm \AA$), as its cool stellar photosphere induces strong CO absorption.

The target distribution of both ends of our $\dot{M}$ interval is consistent with the sample in the \citetalias{connelley18} diagram, where bona fide FUors and FUor-like objects form a tight group and are delineated from other YSOs. While  $\dot{M}$ primarily determines the position of the target in the diagram, $M_*$ also plays a role: a lighter central mass puts a target closer to the FUor cluster given the same $\dot{M}$. This trend agrees with that of $\eta$ contour lines in Fig.~\ref{fig acc_compare}, where a lower central mass requires a lower accretion rate to reach a certain $\eta$ value. For intermediate accretion rates, our model prediction 
lies close to the region occupied by peculiar objects. Although this coincidence may suggest the interpretation of the peculiar objects as star-disk systems with low-accretion rates, the intricacy of their observed spectra likely defies such an easy explanation from our simple model.

Viewing the diagram developmentally, we predict how the CO and metallic line features will change if a YSO undergoes an accretion burst. At the initial stage, the increase of the accretion rate heats up the dust disk internally, which veils both CO and metallic absorption to nearly zero. 
The viscous disk emerges with further increases in $\dot{M}$;
the YSO thus migrates along the colored points as its viscous disk component grows from negligible to dominant, gradually boosting the EWs of both CO and Na+Ca, until the YSO reaches the cluster of very luminous FUors.

Our model does not include CO overtone emission, common to smaller EX Lup-type outbursts \citep[e.g.][]{lorenzetti09,aspin10,kospal11,giannini22} and other accretion disks \citep[e.g.][]{najita96,leeS16}.  This emission is likely generated in a heated disk surface, though a wind origin is possible (see \citealt{pontoppidan11} for analysis of CO fundamental bands), and requires a more sophisticated approach that includes chemistry to incorporate into our synthetic spectra.  However, CO emission {from the passively heated disk} may fill in and weaken some of the CO absorption bands in the spectra, e.g, possibly for V883 Ori (Sect.~\ref{v883ori}).

\subsection{Water vapor absorption in the $H$-band}
\label{water vapor absorption}

\begin{figure*}[ht]
    \centering
    \includegraphics[width=\textwidth]{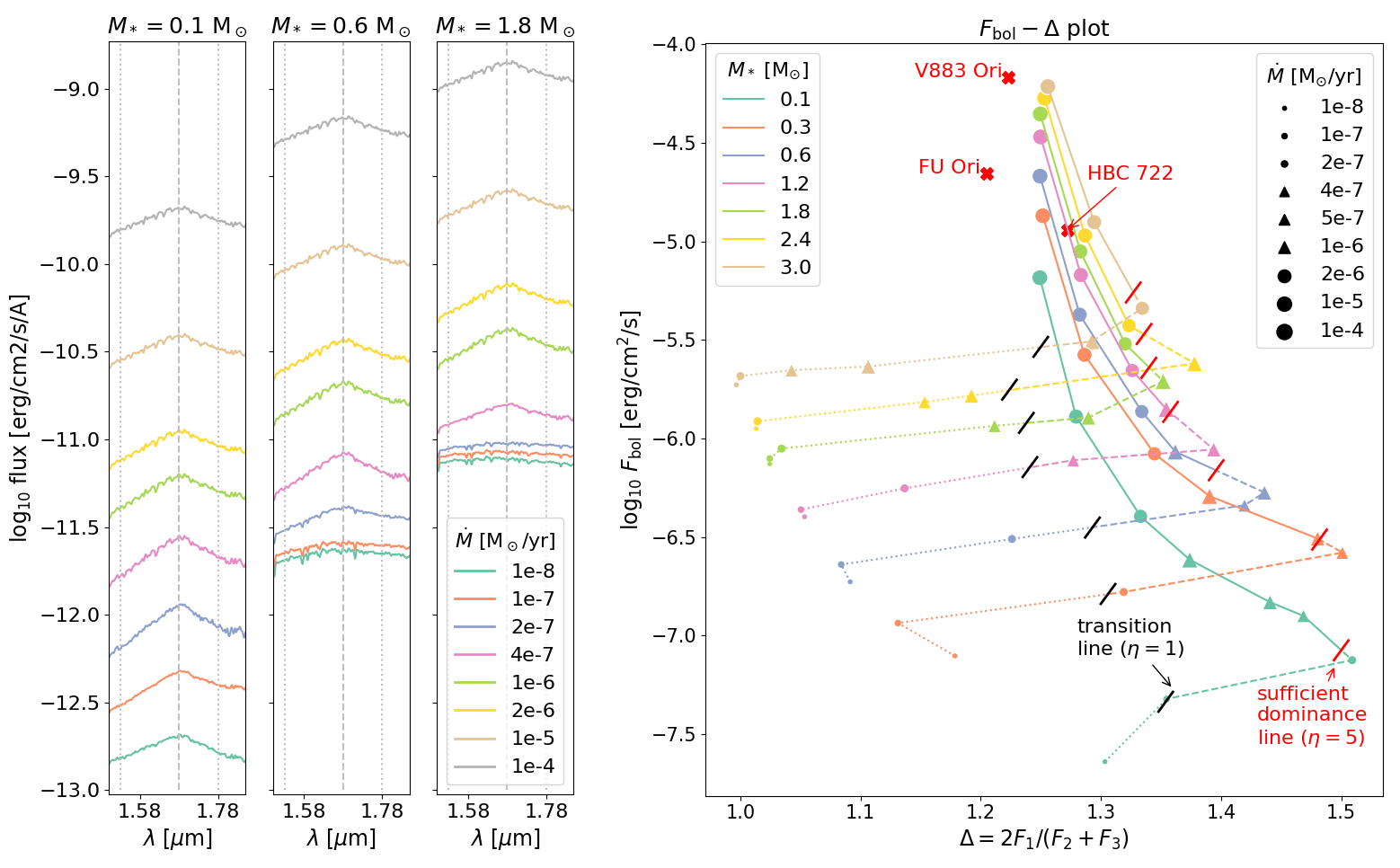}
    \caption{{\it Left three:} Examples of triangle-shaped SEDs at $\lambda = 1.68~\rm \mu m$ for a range of accretion rates and extinction $A_V=10$ mag. The sub-panels show a range of accretion rates for specific stellar masses ($0.1~{\rm M_\odot}, 0.6~{\rm M_\odot}$ and $1.8~{\rm M_\odot}$ respectively). Dashed vertical lines indicate the peak of the triangle $(F_1, 1.68~\rm \mu m)$ and the dotted lines $F_2\ (1.53~\rm \mu m)$ and $F_3\ (1.78~\rm \mu m)$. {\it Right:} Bolometric flux and the color index $\Delta=2F_1/(F_2+F_3)$ of model objects with different masses and accretion rates. The three specific fluxes are measured at wavelengths $\lambda = 1.68~\rm \mu m, 1.53~\mu m, 1.78~\mu m$ respectively. Black and red dashes respectively mark the transition line and the sufficient dominance line, which separate the iso-mass curves into three segments: $\eta<1$ (dotted), $1<\eta<5$ (dashed) and $\eta>5$ (solid). {Names of FUors in Sect.~\ref{fits to data} are indicated and annotated in red.} Extinction is set at $A_V=10$ mag.}
    \label{fig color_index_diagram}
    \vspace{1em}
\end{figure*}




The triangular continuum profile in the $H$-band, shaped by water absorption on both sides, has been observed among substellar objects with low surface gravity and cool photosphere \citep[e.g.,][]{patience12}. Since the viscous disk features low gravity and a wide range of temperature, therefore showing a distinguishing triangular spectral shape, \citetalias{connelley18} used the characteristic to set FUors apart from YSOs without a bright viscous disk.

The presence of the $H$-band triangular shape is thus closely related to the proportion of the viscous disk luminosity to the stellar photospheric luminosity ($\eta$ in Sect.~\ref{viscous disk versus photosphere}). With our model, we predict how the triangular shape of the $H$-band continuum develops when the accretion rate of a YSO grows from quiescence through sufficient dominance of the viscous disk.

We produce a series of synthetic triangular continua of a wide range of $M_*$ and $\dot{M}$, with some examples shown in Fig.~\ref{fig color_index_diagram} ({\it left panels}).
To describe the triangular peak quantitatively, we select three specific fluxes: $F_1$, the peak, at $\lambda = 1.68~\rm \mu m$, and $F_2$ and $F_3$ on both sides at $\lambda = 1.53~\rm \mu m$ and $1.78~\rm \mu m$. Then, we calculate the color index $\Delta$ defined as $\Delta\equiv2F_1/(F_2+F_3)$. The choice of three wavelengths minimizes the influence of extinction so that our analysis remains valid even in the case of heavily embedded YSOs.

We show the bolometric flux of model YSOs versus $\Delta$ in a color-magnitude-diagram style in Fig.~\ref{fig color_index_diagram} ({\it right}). Viewing the figure along iso-mass curves, YSOs with sufficiently high accretion rates are distributed on the steep branch. On the other hand, from the perspective of the viscous disk versus the photosphere, $\dot{M}$ is sufficiently high if the former outshines the latter by a factor of $\eta\sim5$, or the sufficient dominance line defined in Sect.~\ref{viscous disk versus photosphere}. The red dash marks in Fig.~\ref{fig color_index_diagram} show that these two views converge. Therefore, the importance of the sufficient dominance line, besides the upper threshold of magnetopheric accretion (Sect.~\ref{viscous disk versus photosphere}), is that it indicates a transition in trend from rapid development to slow attenuation of the triangular feature as a YSO undergoes an increase in brightness. {FU Ori, V883 Ori and HBC 722 are also presented here, with $\Delta$ measured from their $H$-band spectra and bolometric luminosity adopted from \citetalias{connelley18} but updated with distances in Table \ref{tab fit}.}


While the line where the viscous disk dominates 
accompanies the most pronounced triangles, surpassing the transition line suffices to produce notable peaky profiles. For example, an $M_* = 0.3~{\rm M_\odot}$ target at the $\eta=1$ threshold (black dash in Fig.~\ref{fig color_index_diagram}) manifests a peak as sharp as $\dot{M}=1\times10^{-4}~{\rm M_\odot}$/yr. Therefore, if based on the sole criterion of water vapor band, we would separate low-luminosity FUors from TTSs with the transition line. At low resolution, there may be some degeneracy between FUor-like spectra and brown dwarfs if the luminosity is not well constrained (see discussion in \citetalias{connelley18}), and the triangle may serve as evidence for classification. However, other criteria can distinguish FUors from stars of mass $M_*\lesssim0.1~{\rm M_\odot}$, whose cool, low surface gravity photospheres also exhibit triangular $H$-band continua.

\begin{figure}
    \centering
    \includegraphics[scale=0.52]{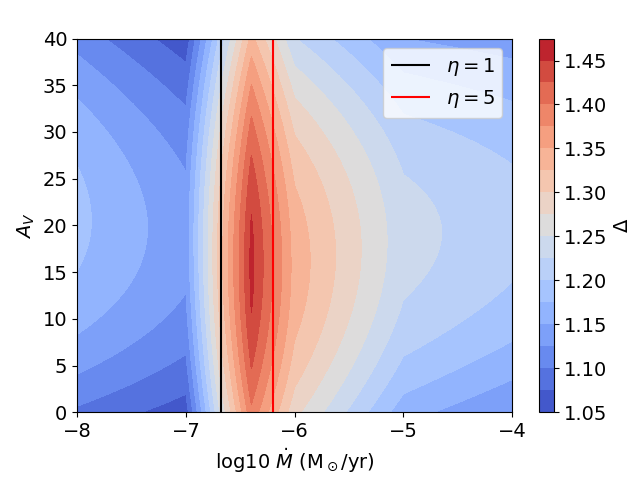}
    \caption{Example ($M=0.3~{\rm M_\odot})$ of contour plot of the color index $\Delta$ as a function of $\dot{M}$ (x-axis) and $A_V$ (y-axis). Black and red vertical lines indicate the transition line and the sufficient dominance line.} 
    \label{fig color_extinction_contour}
\end{figure}

Extinction from the interstellar medium and any protostellar envelope reddens the emission, introducing shifts to measurement of color indices. 
The influence of extinction can be controlled to less than 0.01 in gradient by measuring the ratio of flux at $\lambda = 1.68~\mu$m to that at $1.53~\mu$m and $1.78~\mu$m (Fig.~\ref{fig color_extinction_contour}).
Taking the range of the index across accretion rates as $\sim0.25$, the maximal error is estimated as $err = 0.01/0.25\times100\% = 4\%$ per total extinction. In particular, at $A_V\sim$10--30 mag, where the median half of the \citetalias{connelley18} sample are located, the index is nearly unchanged by extinction. {Objects with low extinction and high accretion rates may present smaller $\Delta$, which explains why FU Ori lies atray in Fig.~\ref{fig color_index_diagram}.} Our quantitative description of the triangular spectral feature can thus undertake homogeneous analysis through various environments of extinction.





\begin{figure*}[!t]
    \centering
    \includegraphics[scale=0.38]{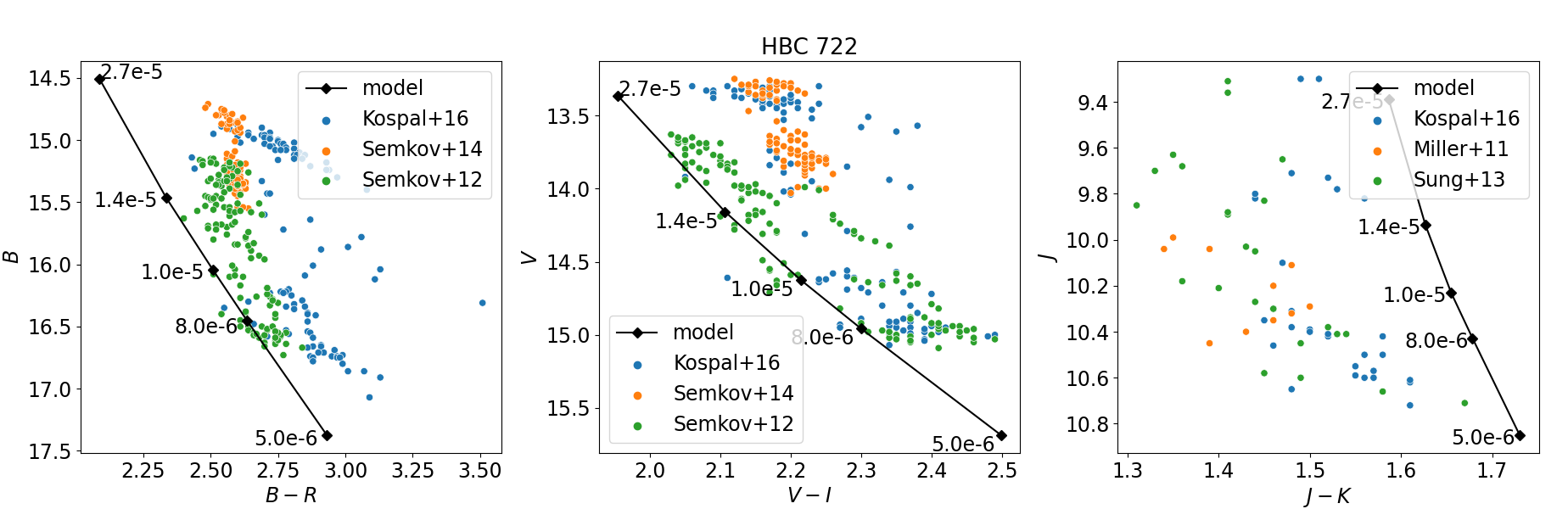}
    \caption{CMD of HBC 722 at optical and near-IR wavelengths. Black line indicates model colors and magnitudes with annotated accretion rates. Colored points show published photometries.}
    \label{fig HBC722_CMD}
    \vspace{1em}
\end{figure*}

\section{Discussion}
\label{discussion}

In the previous section, we calculate models for FUor-like objects across a wide range of parameter space and then develop quantified diagnostics to assess the presence of viscous disks.  In this section, we apply those results to specific observations to infer physical parameters of the FUor-like eruptions.

\subsection{HBC 722: Infrared colors as diagnostics of extinction}

Once the viscous accretion dominates emission, higher accretion rates lead to hotter maximum temperatures in the inner disk.  Cooler regions move to larger radii, increasing in the surface area of the viscous disk.  In our fiducial model, as the accretion rate increases from $5\times10^{-6}$ to $1.5\times10^{-5}$~M$_\odot$/yr, the $V$-band gets $2.1$ mag brighter and $V-I$ gets 0.4 mag bluer, while $J-K$ only gets 0.12 mag bluer.  

As is shown in the CMDs (Fig.~\ref{fig HBC722_CMD}), these color changes are consistent with photometric monitoring of HBC 722 during the initial rise, with photometry from \citet{semkov12} that shows a $V$-band brightness increase of 1.4 mag and $V-I$ getting 0.35 mag bluer (similar to other colors).  In the dip and subsequent rise, similar brightness changes of $\Delta V\sim 1.5$ correspond to 0.2 mag differences in the $V-I$ colors \citep{kospal16}.  These bluer colors were measured and discussed previously \citep{semkov14,baek15,kospal16} and interpreted as an increase in emission from higher temperature material.  Meanwhile, the near-IR colors stays consistent to within $\sim 0.05$ mag in $J-H$ and 0.15 mag in $J-K$ during the brightness changes, with no strong dependence on brightness \citep{miller11,sung13,kospal16}.

The consistency in the near-IR colors from HBC 722 and in FUor models indicate that $JHK$ colors should be reasonably accurate diagnostics of extinction.  However, optical colors are highly dependent on accretion rate, with significant degeneracy between extinction and accretion rate.

\subsection{Timescales for lag of brightening between the mid-IR and the optical}
\begin{figure}
    \centering
    \includegraphics[scale=0.37]{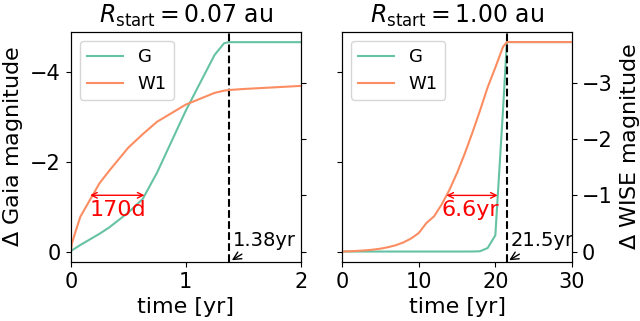}
    \caption{{Lightcurves of WISE $W_1$ and {\it Gaia} $G$ bands with different starting radii ({\it left: }$R_{\rm start}=0.07$~au; {\it right: }$R_{\rm start}=1.00$~au). The two panels share the y-axis. Intervals, marked in red, are measured from $\Delta W_1=-1.00$~mag through $\Delta G=-1.25$~mag. Black dashed lines denote when the propagation reaches the disk inner boundary.}}
    \label{fig lag_lightcurve}
\end{figure}
The multi-wavelength lightcurves of some recent FUor outbursts are thought to imply a lag of several months between the brightness increase in the mid-IR and that in the optical (Gaia17bpi, \citealt{hillenbrand18}; Gaia18dvy, \citealt{szegedi-elek20}; see discussion also in \citealt{fischer22}). An inward propagation process has been proposed, in which some farther regions of the disk brightens before the closer and hotter regions.  We reproduce the required lag with a naive application of our model combined with an analytical estimate of accretion timescales through an alpha disk \citep[see also][]{lee_20_ec53}.

At the outset of the outburst, we impose an immediate {axisymmetric} increase of the accretion rate to $\dot{M}_{\rm fast}=1\times10^{-5}$~M$_\odot/$yr on {a star-disk system with $M_*=0.3$~M$_\odot, \dot{M}_{\rm slow}=1\times10^{-7}$~M$_\odot/$yr, and $R_* = 1.6$~R$_\odot$. We initially let this increase occur only at a narrow annulus of radius $R_{\rm start}$ in the disk, the location of a supposed perturbation that triggers the outburst.} We then let the fast-accretion state propagate {radially} through the disk by following the estimations of  \citet{bellin95}, with an outside-in rate $\sim\alpha c_s$ and an inside-out rate $\sim\alpha c_s H/r$, where $\alpha$ is the viscous efficiency parameter of the $\alpha$-disk model, $c_s$ is the speed of sound, and $H/r$ is the disk scale height divided by the local radius. With $R_{\rm start}=0.07$ au, and setting $c_s=100$~R$_\odot/{\rm yr}$ (corresponding to a temperature of $\sim1000$~K) and $\alpha=0.1, H/r=0.3$ for the inner disk, we reproduce a lag of $\sim200$ days between the brightening of the WISE $W_1$ band and the {\it Gaia} $G$ band, shown in Fig.~\ref{fig lag_lightcurve} ({\it left}), with the optical brightening process lasting $\sim500$ days. {The $W_1$ band continues to rise slowly after the propagation reaches the disk inner boundary, eventually reaching maximal brightness with $\Delta W_1/\Delta G$=4/5}. 

This model is in good agreement with observed timescales for the starting radius $R_{\rm start} = 0.07$~au, corresponding to a local temperature $T=2400$ K in the fast-accretion state and is fairly close to the central star. {A farther $R_{\rm start} = $1.00~au leads to a much longer (6.6~yr) interval and very different rising slopes between $W_1$ and $G$ bands (Fig.~\ref{fig lag_lightcurve}, {\it right})}. We adopt a large $\alpha$ value, which speeds up propagation.  Setting $\alpha=0.01$ instead would extend both the brightening timescale and the lag by a factor of 10. We do not consider the timescale of the rising process of $\dot{M}$ from the quiescent to the outburst state, which means the mid-IR brightening could start still earlier than the optical. These two factors imply that the real $R_{\rm start}$ may be even smaller. Our estimation suggests that if such FUor outbursts with lags are triggered by a local perturbation, the location of this perturbation is expected to be close to the star.

\subsection{FUor-like objects among the youngest protostars}

Early in the stellar growth, the star has a low mass but high accretion rate \citep[e.g.,][]{kunitomo17,arturdelavillarmois19}.  In the simple growth models of \citet{fischer17hops}, accretion rates smoothly decline with time, staying above $3\times10^{-6}~{\rm M_\odot}/$yr for $\sim5\times10^4$~yr.  For these very young objects, the spectrum may appear FUor-like because the inner disk is viscously heated, despite not having any accretion burst.  
The high rate of FUor-like objects reported from near-IR spectra by \citet{Connelley10} may have some contribution from high accretion rates onto protostars that are still early in their mass assembly {and not necessarily in a large burst.}  Although very young photospheres are challenging to measure \citep[see, e.g.,][]{laos21,yoon22}, when detected the presence of deep CO absorption may indicate a viscous disk rather than a true FUor outburst. Our models are useful to diagnose the parameter space of these youngest FUor-like objects to better understand their physical properties.

Before the central star assembles most of its material, the low stellar mass and density (hence relatively large $R_{\rm in}$) entails a cooler viscous disk if $\dot{M}$ is constant. On the other hand, the photosphere rapidly brightens as the star gains mass and expands, reaching peak luminosity at $\sim6\times10^{4}$~yr, after which the isothermal contraction dims the photosphere. Therefore, higher accretion rates are needed for the viscous disk to outshine the photosphere. Indeed, for a final central mass of $M_*=0.5~{\rm M_\odot}$, the typical $\dot{M}$ of the transition line and the sufficient dominance line are (1--2)$\times10^{-6}~{\rm M_\odot}$/yr and (3--6)$\times10^{-6}~{\rm M_\odot}$/yr respectively at an age younger than 0.1 Myr; in the most extreme case, when the YSO has just finished the deuterium-burning phase, the transition line and sufficient dominance line can be as high as $3.6\times10^{-6}~{\rm M_\odot}$/yr and $9.5\times10^{-6}~{\rm M_\odot}$/yr.

We describe three example protostars to describe what may be expected for infrared spectroscopy of very young protostars, if their photosphere is detectable directly or in scattered light.

The protostar IRAS 15398-3359 has a central mass of 0.007 ${\rm M_\odot}$ \citep{okoda18} and a present-day total bolometric luminosity of $1.8$~${\rm L_\odot}$ \citep{jorgensen13}. Our models require $\dot{M} \sim 10^{-5}~{\rm M_\odot}$/yr to reproduce this measured luminosity; the near-IR spectra, with both viscous-disk features and photospheric absorption, if detected would likely appear similar to those of T Tauri star-disk systems around the transition line, perhaps including strong CO emission. The parameter space of accretion rate where IRAS 15398-3359 would have optical and near-IR emission dominated by a viscous disk is similar to the parameter space for growing protoplanets, as described by \citet{zhu15}.

The protostar EC 53, also known as V371 Ser, has a mass of 0.3 ${\rm M_\odot}$ \citep{leeS_20} and a variable luminosity between 6--20~${\rm L_\odot}$ \citep[e.g.,][]{baek20}. Assuming that the star finally grows to $M_* = 0.5\ {\rm M_\odot}$, we apply the \cite{kunitomo17} model with a stellar age of $t = 0.09$~Myr, $R_* = 3.8\ {\rm R_\odot}$ and $T_* = 3500$~K. Accretion rates in a range of (0.3--1.4)$\times10^{-5}~{\rm M_\odot}/$yr would result in near-IR spectra that spans the transition region through sufficient dominance of the viscous disk. The high end should closely resemble FUors, with prominent CO absorption, the $H$-band triangular continuum and few traces of the stellar photosphere, whereas the quiescent state should show mixed characteristics.

The FUor-like object IRAS 16316-1540 \citep{yoon21}, with a very low luminosity of $\sim 2$~L$_\odot$, is difficult to explain using this viscous disk model.  The $H$ band spectrum lacks the strong triangular shape that is characteristic of FUors and more generally of sources dominated by low temperature emission. To reproduce the spectrum adequately, we have to cut off the gas disk where $T\lesssim3000$~K to avoid a prominent triangular shape, while maintaining the dust disk to fit the bright $K$-band continuum. Further study is needed to understand how the photospheric absorption lines appear double-peaked, consistent with expectations from a heated viscous disk, without producing the triangular shape of the $H$-band.

\subsection{The hottest inner disks}

\cite{hillenbrand19} reported PTF 14jg, a previously faint source that underwent a peculiar eruption in 2013. Multi-wavelength photometric and spectroscopic analyses lend evidence to identifying the event as an FUor-like outburst, but the source exhibits lines unusually hot ($\sim$11,000--15,000~K) for a viscous disk and no CO absorption. It is therefore intriguing what the spectrum of an FUor would look like at extremely high accretion rates.

Our model shows that for a disk with $\dot{M}=1\times10^{-3}~{\rm M_\odot}$/yr and a maximal disk temperature $T_{\rm max}=21,500$~K, the strong CO absorption still lingers, with an EW of $20.5\ \rm\AA$. Our interpretation is that although the $T\sim10^4$~K regions do not produce CO band profiles characteristic of FUors and instead tend to veil possible CO absorption, the high $T_{\rm max}$ also implies larger areas of moderate temperature capable of generating such CO lines (see Sect.~\ref{accretion rate}).

On the other hand, a very rapidly accreting disk with a small outer radius $R_{\rm out}$ could maintain high temperature throughout the disk and thus lack CO absorption.  The EW of CO is reduced to less than $1\ \rm\AA$ when the outer radius of this hot disk is set to $R_{\rm out} < 0.08$ AU. 
In this case, the outer boundary of the disk $T_{\rm out} > 6000$~K, thus devoid of the dust disk component. This scenario, however, is challenged by the deficient infrared excess and physical plausibility.




\subsection{At the boundaries of FUor-like objects}

Many eruptive YSOs show spectral and photometric properties that do not completely agree with those usually associated with FUors  \citep[e.g.,][]{contreras17,guo21}.  If this class of misfits, or {\it peculiar} objects \citepalias{connelley18}, has an archetype object, it would be V1647 Ori, a source that has spectral characteristics that appear to combine a viscously heated disk with magnetospheric accretion.

YSOs with accretion rates from $3\times10^{-7}$ to $10^{-6}$ M$_\odot$ yr$^{-1}$ (see Fig.~\ref{fig acc_compare}) are at the edge of where viscous or photospheric emission dominates in the infrared.  Such objects may have emission from both the photosphere and the disk, unless the innermost disk completely envelops the star; the ratio of disk to photosphere emission should decrease to shorter wavelengths.  Changes in accretion rate of a factor of a few could alter whether the viscous disk or stellar photosphere and magnetospheric accretion dominate.  
Indeed, many of the peculiar objects, including EC 53 discussed above, have luminosities that are consistent with accretion rates in this range.

Besides showing mixed emission features that challenge stellar classification, this range is also strongly mass-dependent (Fig.~\ref{fig acc_compare}, Sect.~\ref{viscous disk versus photosphere}), which has implications on what should be called an FUor event. If we maintain the criteria throughout this work, namely, a decades-long outburst with spectrum characteristic of a viscously heated accretion disk, a $0.5~{\rm M_\odot}$ YSO accreting at $-6.5<\log \dot{M}<-6.0$ for longer than 10 years will fulfill the definition, but a $2.4~{\rm M_\odot}$ YSO with identical accretion rates may not because its viscous disk does not sufficiently dominate emission. This might explain the case of VVVv270, a $\sim2~{\rm M_\odot}$ object ({estimated with a large uncertainty} from SED modelling) that has a very long outburst but CO overtone bands in emission \citep{contreras17,guo20}.  However, in both cases ($0.5~{\rm M_\odot}$ and $2.4~{\rm M_\odot}$) the same amount of material is accreted. 

We thus speculate that the accretion rate is not large enough for the viscous disk to dominate emission. In fact, many VVV outbursts in \citet{guo21} show long-duration outbursts but are dominated by magnetospheric accretion and show CO in emission rather than absorption.
The physical mechanisms that lead to the outburst might be very similar, but our current spectroscopic criteria for FUors would place  them in different categories. Another complexity is that even though a YSO exhibits an outburst with viscous disk characteristics, the high state may not last for $\gtrsim10$ years \citep[e.g., VVVv322 in ][]{contreras17}. In this case, the central star does not accrete a significant fraction of the stellar mass during the outburst as classical FUors do. If we classify such YSOs as FUors, the role of FUor outbursts in stellar mass assembly will become more heterogeneous. 


\section{Conclusions}
\label{conclusions}
We have presented an FUor-type disk model that includes four emission components: the viscously heated disk, the magnetospheric accretion, the stellar photosphere and the passively heated dust disk, with parameters constrained by standard PMS evolutionary models. Our model agrees with observed FUor photometries and spectra.

In our exploration of the parameter space, we show that $M_*$ and $\dot{M}$ are vital parameters that profoundly influence the shape of the SED and spectral lines. Parameters such as $R_*$, $R_{\rm in}$ also play important roles, but it is difficult to constrain them accurately with our models alone: the standard PMS model treatment leads to higher $T_{\rm max}$ estimation of FUors, and the dipole magnetic field assumption may induce inaccurate estimation of $R_{\rm in}$. {We also do not include any possible CO emission, which could arise from the surface of the passively heated disk and may help to fill in the CO absorption bands.}  These limitations, along with other simplification in the current work, invites more detailed theories (e.g., analysis on the material, thermal and magnetic star-disk interaction near the disk inner boundary) and external observational evidence to better understand FUors in the parameter space.

We show physical and observational significance of the $M_*$- and $\dot{M}$-dependent value of $\eta$, the ratio between emission from the viscously heated disk to the emission from the central star. The $\eta=1$ transition line and the $\eta=5$ sufficient dominance line respectively delineate where the viscous disk starts to outshine the star and to dominate emission, and the sufficient dominance line concurs with the disappearance of the magnetospheric accretion. In terms of observational FUor characteristics, the two lines indicate potentially fruitful regions in search of new FUor-like candidates in the optical and mid-IR CMDs, and also mark the development of strong CO absorption and triangular $H$-band continuum profile.

 Our FUor-type disk models are physically similar to previous studies \citep{rodriguez22,hillenbrand22}, but we incorporate the magnetospheric emission component and standard PMS models. Our work explores the broad parameter space thoroughly to show detailed relations between physical parameters and observational traits and to evaluate when a disk might be identified as FUor-like.
Our models provide guidance for optical and infrared emission of FUor-type objects and more generally of YSOs showing evidence of viscous disk emission, along with implications on stellar mass assembly. Future work may extend to analysis on other physical aspects of FUor-type objects and application to a larger domain of viscous-disk related phenomena.

\acknowledgements

{We thank the anonymous referee for providing insightful comments and suggestions that helped to improve this paper.}
GJH, DJ, and JL thank the JCMT Transient Team for providing motivation for this work.  GJH and DJ thank Will Fischer, Lynne Hillenbrand, Agnes Kospal, and Mike Dunham for extensive discussions during the preparation of a Protostars and Planets VII chapter that helped to frame some of the applications.  GJH and HL thank Lee Hartmann for early questions.  We also thank Nienke van der Marel for contributions to the WHT spectrum of HBC 722.  

HL and GJH are supported by general grants 12173003 and 11773002 awarded by the Natural Science Foundation of China.  D. Johnstone is supported by the National Research Council Canada and an NSERC Discovery Grant. J.-E. Lee and C. Contreras Pena are supported by the National Research Foundation of Korea (NRF) grant funded by the Korea government (MSIT) (grant number 2021R1A2C1011718).

These models support the interpretation of data obtained in programs awarded on Gemini Observatory through Korea and Canada, on NASA/IRTF, and on Palomar awarded by the Telescope Access Program in China.  This publication makes use of data products from the Wide-field Infrared Survey Explorer, which is a joint project of the University of California, Los Angeles, and the Jet Propulsion Laboratory/California Institute of Technology, funded by the National Aeronautics and Space Administration.  This research has made use of the Spanish Virtual Observatory (\url{https://svo.cab.inta-csic.es}) project funded by MCIN/AEI/10.13039/501100011033/ through grant PID2020-112949GB-I00.  This work has made use of data from the European Space Agency (ESA) mission
{\it Gaia} (\url{https://www.cosmos.esa.int/gaia}), processed by the {\it Gaia}
Data Processing and Analysis Consortium (DPAC,
\url{https://www.cosmos.esa.int/web/gaia/dpac/consortium}). Funding for the DPAC
has been provided by national institutions, in particular the institutions
participating in the {\it Gaia} Multilateral Agreement.


\appendix

\section{Disk gravity profile}
For the model spectra, we have assumed that the gravity is constant, adopting $\log g = 1.5$ (see Sect.~\ref{subsec viscous disk}; hereafter uniform gravity model). This choice is based on empirical fit of line strengths of FU Ori. In comparison, more detailed 
FUor disk models consider variation of the gravity both radially and vertically. Calculations by \cite{Calvet91} link $g = g(r,z)$ to the radial and vertical distance $r$ and $z$ as well as the stellar mass $M_*$, stellar radius $R_*$ and disk height $H$:
\begin{equation}
    g(r,z) = \frac{GM_*H}{r^3}\frac{1+z/H}{[1+(z+H)^2/r^2]^\frac{3}{2}}
\label{eq gravity}
\end{equation}
where $H$ is assumed as $H = H_0(r/R_*)^{9/8}$, with $H_0 = 0.1R_*$.
\begin{figure}[ht]
    \centering
    \centering
    \includegraphics[scale=0.45]{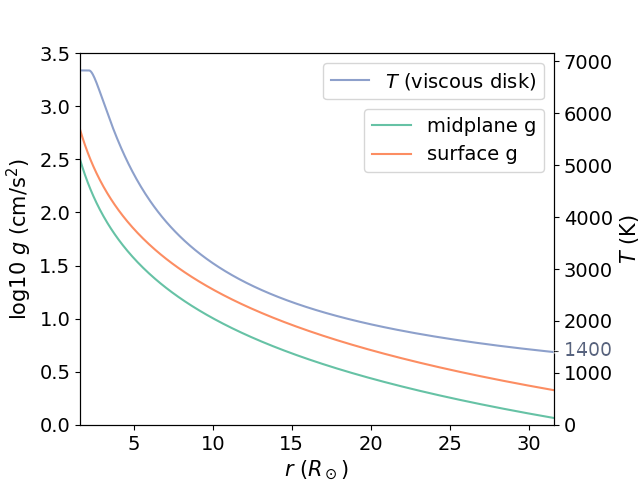}
    \caption{The midplane and surface gravity versus radial distance in the viscous disk region. The temperature profile is also included. Dusty regions outside the range of this plot, where we assume blackbody emission, is not affected by gravity. Fiducial disk properties are adopted.}
    \label{fig gravity}
\end{figure}
We thus construct a modified version of our disk model which considers the radial gravity distribution (varying gravity model). We adopt the midplane gravity value ($z=0$) of a given annulus and ignore vertical variation, since the viscous heating mainly occurs in deeper layers of the disk and that the vertical change of gravity does not exceed a factor of 1.5, negligible compared with the range of radial variation of more than 2 orders of magnitude. 
\begin{figure*}
    \centering
    \includegraphics[width=\textwidth]{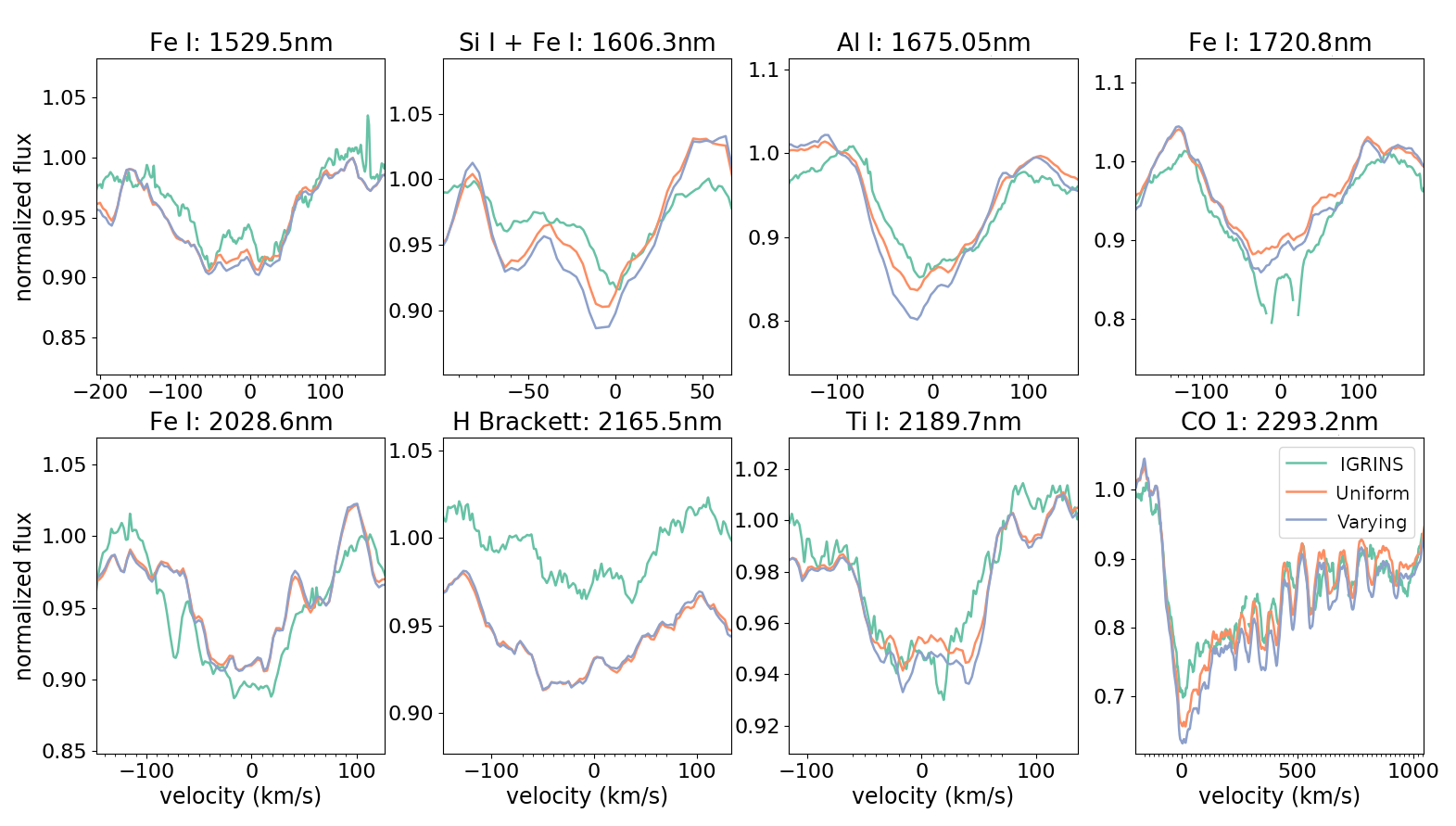}
    \includegraphics[width=\textwidth]{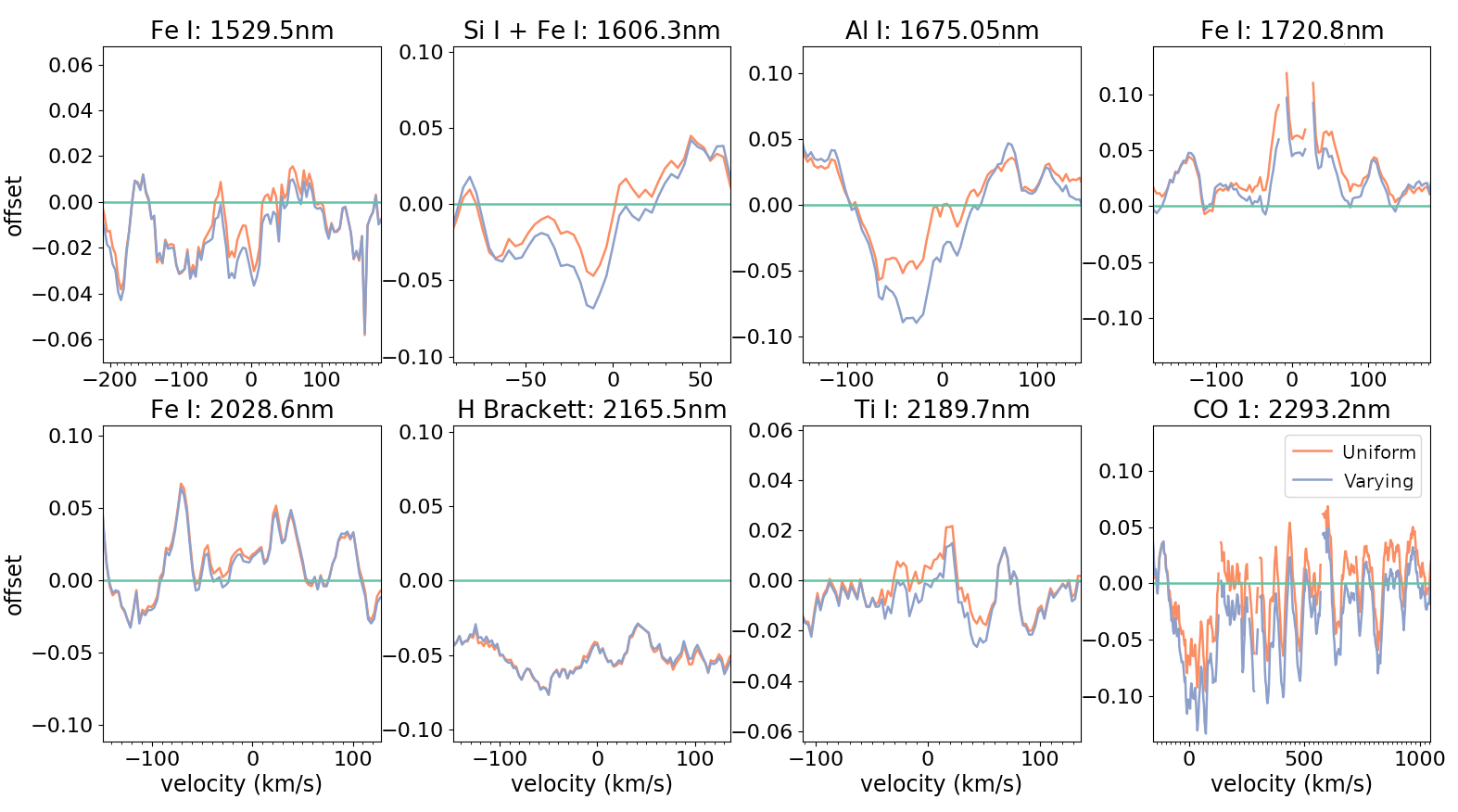}
    \caption{Model and observed spectral lines of FU Ori ({\it top}) and differences between the models and observations ({\it bottom}), for the uniform gravity model (orange lines) and the varying gravity model (blue lines).}
    \label{fig FU_Ori_lines_g}
\end{figure*}
The varying gravity model yields slightly stronger lines compared with the uniform gravity model (Fig.~\ref{fig FU_Ori_lines_g}). This can be explained by the power-law relationship $T_{\rm eff} \propto g_{\rm eff}^\beta$ \citep[e.g.,][]{vonzeipel24,lara11}. The varying gravity model predicts lower average $g_{\rm eff}$; where $T_{\rm eff}$ is several thousand Kelvin, higher average gravity translates into higher $T_{\rm eff}$ and thus weaker features. The widths of lines predicted by two models do not differ because Doppler broadening in the viscous disk dominates over pressure broadening.

Applying the varying gravity model to FU Ori, we arrive at good photometric fit results with similar parameters as the uniform gravity model. For absorption lines, while the uniform and varying gravity models agree with each other in terms of line shape and width, the strength of some lines is slightly more pronounced in the varying gravity model, e.g., \ion{Al}{1}\ ($1.675~\rm \mu m$) and the first CO overtone band (2.292--$2.320~\rm \mu m$).

\begin{deluxetable}{cccccccc}[!h]
    \tablecaption{Comparison between uniform and varying gravity model best fits for FU Ori.}
    \tablehead{\colhead{Model} & \colhead{$M_*$} & \colhead{$\dot{M}$} & \colhead{$i$} & \colhead{$T_0$} & \colhead{$A_V$} & \colhead{$\chi^2$} \\ 
    \colhead{} & \colhead{(${\rm M_\odot}$)} & \colhead{(${\rm M_\odot}$/yr)} & \colhead{} &  \colhead{(K)} & \colhead{(mag)} & \colhead{} } 
    \startdata
    Uniform & 0.6 & $2.8\times10^{-5}$ & $37^{\circ}$ & 4000 & 2.1 & 21.6\\
    Varying & 0.6 & $2.9\times10^{-5}$ & $37^{\circ}$ & 4000 & 2.2 & 21.4\\
    \enddata
\vspace{-2.5em}
\end{deluxetable}

In view of the adequately consistent continuum fit of the uniform and varying gravity models and the marginally superior performance of the uniform gravity model in reproducing FU Ori line depths, we adopt the simpler setting of uniform gravity. Still, both models leave imperfections in line shapes and depths. The dissimilarity may stem from physical and chemical differences between the photosphere, which the BT-Settl synthetic spectra are based on, and the FUor disk environment. More elaborate models of the FUor gravity profile will become testable if the intrinsic mismatch between theoretical and observational spectra is better answered.

\bibliographystyle{aasjournal}
\bibliography{ms.bib}
\end{document}